\documentclass{emulateapj}
\usepackage{graphicx}
\usepackage{amssymb}
\usepackage{subfigure}
\newcommand{\degree}{\ensuremath{^\circ}}
\usepackage{natbib}
\usepackage{comment}
\usepackage{enumerate}
\usepackage[T1]{fontenc}
\usepackage{pslatex}

\shorttitle{NGC 4330}
\shortauthors{Abramson et al.}
\begin{document}
\title{Caught in the Act: Strong, Active Ram Pressure Stripping in Virgo Cluster Spiral NGC 4330}

\author{Anne Abramson \& Jeffrey D. P. Kenney}
\affil{Department of Astronomy, Yale University, P.O. Box 208101, New Haven, CT 06520, U.S.A.}
\email{anne.abramson@yale.edu}
\author{Hugh H. Crowl}
\affil{Department of Astronomy, Columbia University, 550 West 120th Street, New York, NY 10027, U.S.A.}
\author{Aeree Chung}
\affil{Department of Astronomy, Yonsei University, 134 Shinchonding, Seodaemungu, Seoul 120-749, Korea} 
\affil{Smithsonian Astrophysical Observatory, 60 Garden St., Cambridge, MA 02138, USA}
\author{J. H. van Gorkom}
\affil{Department of Astronomy, Columbia University, 550 West 120th Street, New York, NY 10027, U.S.A.}
\author{Bernd Vollmer}
\affil{Observatoire astronomique de Strasbourg, 11 rue de l'universite, 67000 Strasbourg, France}
\and
 \author{David Schiminovich}
 \affil{Department of Astronomy, Columbia University, 550 West 120th Street, New York, NY 10027, U.S.A.}

\begin{abstract}

We present a multi-wavelength study of NGC 4330, a highly-inclined spiral galaxy in the Virgo Cluster which is a clear example of strong, ongoing ICM-ISM ram pressure stripping.
The HI has been removed from well within the undisturbed old stellar disk, to 50\% - 65\% of R$_{25}$.
Multi-wavelength data
(WIYN\footnote{The WIYN Observatory is a joint facility of the University of Wisconsin-Madison, Indiana University, Yale University, and the National Optical Astronomy Observatory. }
BVR-H$\alpha$, VLA\footnote{The VLA is operated by the National Radio Astronomy Observatory, which is a facility of the National Science Foundation (NSF), operated under cooperative agreement by Associated Universities, Inc.} 21-cm HI and radio continuum, and GALEX\footnote{This work is based in part on observations made with the NASA Galaxy Evolution Explorer.
GALEX is operated for NASA by the California Institute of Technology under NASA contract NAS5-98034.} NUV and FUV)
reveal several one-sided extraplanar features likely caused by ram pressure
at an intermediate disk-wind angle.  At the leading edge of the interaction,
the H$\alpha$ and dust extinction curve sharply out of the disk
in a remarkable and distinctive ``upturn" feature
that may be generally useful as a diagnostic indicator of active ram pressure.
On the trailing side, the ISM is stretched out in a long tail which contains
10\% of the galaxy's total HI emission, 6 - 9\% of its NUV-FUV emission, but only 2\% of the H$\alpha$.  
The centroid of the HI tail is downwind of the UV/H$\alpha$ tail,
suggesting that the ICM wind has shifted most of the ISM downwind
over the course of the past 10 - 300 Myr.  Along the major axis, the disk is highly asymmetric in the UV, but more symmetric in H$\alpha$ and HI, also implying recent changes in the distributions of gas and star formation.
The UV-optical colors indicate very different star formation histories for the leading and trailing sides of the galaxy.
On the leading side, a strong gradient in the UV-optical colors of the gas-stripped disk suggests that it has taken 200-400 Myr to strip the gas from a radius of $>$8 to 5 kpc, but on the trailing side there is no age gradient.
All our data suggest a scenario in which NGC~4330 is
falling into cluster center for first time
and has experienced a
significant increase in ram pressure over the last 200-400 Myr.

\end{abstract}

\section{Introduction}
There is ample evidence that cluster galaxies are strongly affected by their environment over time.  Denser environments in the nearby universe have a higher fraction of elliptical and S0 galaxies (the density-morphology relationship of Dressler 1980) and the spirals are systematically HI-deficient (Haynes \& Giovanelli 1986, Solanes et al. 2001), clusters at z$>$0.1 have a higher fraction of blue, star-forming galaxies than nearby clusters (Butcher \& Oemler 1978, 1984), and at z$\sim$0.5, the fraction of spirals increases while the fraction of S0's decreases relative to that in the local universe (Dressler et al. 1997).  This suggests that one or more cluster processes act to transform spiral galaxies into earlier-type spirals and S0's.  Many effects have been presented as possible contributors, including galaxy-galaxy harrassment (Bekki 1999, Moore et al. 1998), galaxy-galaxy collisions (Kenney et al. 2008), galaxy interactions with the cluster potential (Moore et al. 1999), ram pressure stripping (Gunn \& Gott 1972, Abadi et al. 1999, Poggianti et al. 1999), turbulent viscous stripping (Nulsen, 1982), and starvation (Larson et al. 1980).  There remains significant debate about the relative importance of these processes.  

Several recent studies of clusters out to a redshift of z$\sim$0.8 support the importance of ICM-ISM interactions as a driver of cluster galaxy evolution.  Many studies have found cluster populations of disky, quiescent galaxies with structure similar to spiral galaxies but with little or no star formation (e.g., van den Bergh 1976, Balogh et al. 1998, van der Wel et al. 2010).  Such galaxies are consistent with gas stripping.  Moran et al. (2007) present evidence from two z$\sim$0.5 clusters that there are multiple pathways for star formation to be quenched in spirals, depending on the properties of the cluster, with ICM-ISM stripping emerging as a major contributing factor in
situations where the ICM is dense enough for stripping to be effective.  From a spectral analysis of galaxies in various environments at z = 0.4 - 0.8, Poggianti et al. (2009) find that that clusters are the preferred location of k+a galaxies. Such galaxies have had their star formation was quenched between 5$\times10^7$ and 1.5$\times10^9$ years
prior to observation, making them a likely transition phase between actively star-forming spirals and passively evolving S0 and Sa galaxies. Since the star formation
quenching efficiency increases with cluster velocity dispersion, Poggianti et al. (2009) propose this transformation is likely the result of ICM-ISM interactions.  These studies document how key galaxy properties differ for large samples of cluster galaxies, but do not by themselves offer direct evidence of the mechanism transforming cluster galaxies.


In order to understand the impact of ram pressure stripping on galaxy evolution, we need to be able to identify galaxies which are clearly experiencing stripping or did at some point in the past, and determine which of a given galaxy's characteristics are due to ram pressure stripping.  In particular, we are interested in finding out how the multi-phase ISM behaves during stripping, how long it takes to strip a galaxy, and how ram pressure and the associated stripping affects the rate and distribution of star formation.  Through detailed studies of the best cases of ram pressure stripping in the nearby Virgo cluster, we hope to identify key parameters of the ram pressure stripping interaction, such as the current strength of ram pressure, the angle between the ICM wind and the disk, the time evolution of these quantities, and the evolutionary stage of the stripping event.  If we can quantify key stripping parameters for enough galaxies, we can understand the impact of stripping on galaxy evolution in different environments.    

There have been many ram pressure stripping simulations over the past few years (e.g., Quilis et al. 2000, Schulz \& Struck 2001, Roediger \& Bruggen 2006, Tonnesen et al. 2007, Tonnesen \& Bryan 2008, 2009, 2010, Kronberger et al. 2008,  Vollmer 2009, Jachym et al. 2009; also see review by E. Roediger 2009).  They are all in general agreement that in most situations, ram pressure stripping is well-approximated by the simple Gunn \& Gott (1972) formula, which fails only in cases of nearly edge-on stripping and short, impulsive episodes of ram pressure (Jachym et al. 2009).  All of the simulations predict that the gas stripped from the disk will form a tail downwind of the galaxy, and all of the groups predict a lack of star formation in the stripped outer disks.  Kronberger et al. (2008) and Tonnesen \& Bryan (2009) both predict enhanced star formation rates due to ICM pressure in certain cases.  Progress has been made in modeling the effects of the stripping process on the multi-phase ISM.  For example, Tonnesen \& Bryan (2009, 2010) model the effects of ram pressure on a clumpy ISM with self-gravity and cooling, and also account for the effects of heating.  However, more work remains to be done.  Many complicated physical processes relevant to the ISM, particularly star formation, are simulated using simple prescriptions, and observations are necessary to determine what actually happens. 

There have also been spectacular examples of individual galaxies being stripped in rich clusters between z$\sim$0.2 and 0.5 (Owen et al. 2006, Cortese et al. 2007, Sun et al. 2007), but at such distances we lack the resolution to see what really happens.  The Virgo cluster is the nearest moderately rich cluster, and provides the best resolution to observe cluster processes at work.  For this reason, we have carried out the VIVA (VLA Imaging of Virgo spirals in Atomic gas) survey, which presents high-quality HI imaging of 53 spirals in the Virgo cluster (Chung et al. 2009).  Using data from the VIVA survey, Chung et al. (2007) noted the presence of HI tails in at least 25\% of large spiral galaxies, a total of 7 galaxies, near the cluster's virial radius, providing striking evidence of galaxies actively losing their gas as they plummet towards the cluster center.  One of these galaxies was NGC 4330, and we have followed up with a detailed multi-wavelength study.

In this paper, we show that NGC 4330 is one of the best examples of ram pressure stripping actively transforming a spiral galaxy in the nearby Virgo cluster.  There is a wealth of evidence that it is being actively stripped, and several interesting phenomena can be clearly seen, some for the first time.  In this multi-wavelength study, we examine the effects of stripping on the galaxy's current star formation, recent star formation history, and the stripping of the HI and dust components of its ISM.  In Section \ref{thegal}, we introduce the galaxy NGC 4330.  In Section \ref{observations}, we describe the observations taken in optical broad-band and H$\alpha$ narrow-band filters, HI, and the UV.  In Section \ref{results}, we describe the galaxy's stellar distribution and optical dust extinction in the BVR bands, the HI distribution and kinematics, and the H$\alpha$ and UV distributions.  In Section \ref{discussion} we discuss the interpretation of the galaxy's properties in the context of an ongoing ICM-ISM interaction.  In Section \ref{conclusion}, we summarize our findings.  Throughout this paper, we assume a distance to the Virgo Cluster of 16 Mpc (Yasuda et al. 1997).

\section{The Galaxy NGC 4330}
\label{thegal}

\begin{figure}[h] 
   \centering
   \includegraphics[width=3.5in]{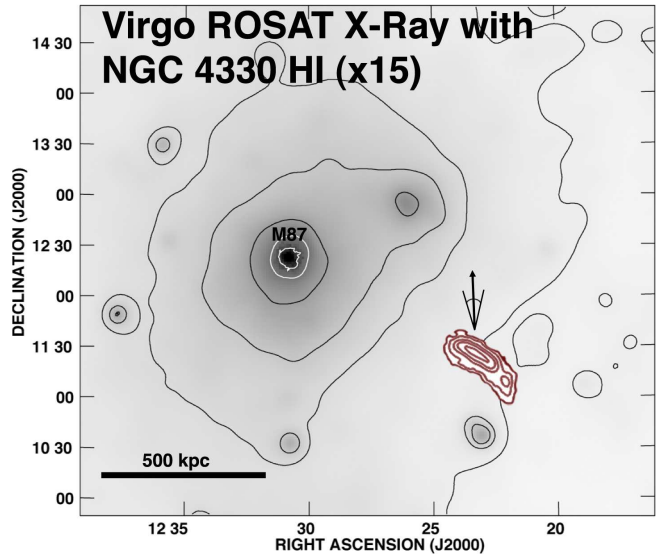} 
   \caption{NGC 4330 HI contours (red) on a ROSAT X-ray (B{\"o}hringer et al. 1994) greyscale and contour map (black) of the Virgo cluster.  The galaxy has been enlarged by a factor of 15.  Note that the tail roughly points away from M87, indicating that the galaxy is probably falling in towards the cluster core.  The arrows indicate the possible range in direction of motion, as inferred from the HI tail angle, the location of the galaxy's radio deficit region, and the position angles of dust features (see also Figure \ref{radiodef} and Section \ref{windangle}).}
   \label{hionrosat}
\end{figure}

NGC 4330 is a late-type, small-bulge, highly-inclined spiral with a maximum rotation speed of $\sim$ 180
	km/s, a luminosity of $\sim 0.3 L_*$, and an HI mass of $4.5\times10^{8}$\ M$_\odot$ (Chung et al. 2009).  It is located at a projected distance of 
	$2\degree$ (600 kpc) from M87 (see Figure \ref{hionrosat}), in a region near the Virgo cluster's virial radius 
	where many galaxies are HI deficient and some have HI gas tails
	likely resulting from ICM ram pressure (Chung et al. 2007).  NGC 4330
	fulfills both of these criteria, making it a strong candidate
	for ongoing ram pressure stripping.  This galaxy was initially noticed in GALEX UV images because of its UV-bright tail.  Chung et al. (2007) find that this galaxy possesses an
	HI tail that extends to the south of the disk plane on
	the western side, pointing roughly away from M87.  Using the assumptions of Gunn \& Gott (1972), Chung et al. (2007) estimate that the ICM ram pressure on the galaxy at this location in the cluster is sufficient to cause the observed stripping of HI.  The tail direction suggests that NGC 4330 is currently falling in towards the cluster core, and the presence of a dense gas tail strongly suggests active stripping.  The galaxy's HI deficiency parameter is 0.8 (Chung et
	al. 2007), meaning that its HI mass is only 15\% of that expected for a field spiral of similar size.  NGC 4330 has a heliocentric
	velocity of 1569 km/s, which exceeds the Virgo Cluster's mean heliocentric systemic velocity of $\sim$1100 km s$^{-1}$.  This indicates that NGC 4330 is likely moving roughly towards the core of the cluster from the front.  With an inclination of $84\degree$, NGC 4330 is nearly
	edge-on.  This makes it an excellent galaxy in which to
	examine the effects of ICM pressure on the galaxy's recent star formation and HI gas morphology, since we can 
	distinguish between planar and extraplanar gas.   
	
\begin{figure}[htbp] 
   \centering
   \includegraphics[width=3.5in]{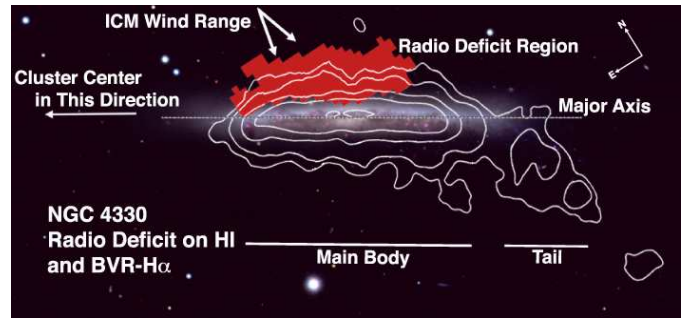} 
   \caption{BVR-H$\alpha$ image of NGC 4330 with radio deficit region (red), HI C-array contours (white), with BVR and H$\alpha$ (pink).  Pixels with a ratio of observed-to-modeled 20 cm radio continuum flux density $<50\%$ are said to have a radio deficit (see Murphy et al. 2009).}
   \label{radiodef}
\end{figure}

NGC 4330 displays a local radio continuum deficit (Figure \ref{radiodef}), which is believed to be a tracer of ram pressure (Murphy et al. 2009).  In a sample of Virgo spiral galaxies examined by Murphy et al. (2009), NGC 4330 had the strongest local radio deficit, suggesting that it is experiencing strong, active ram pressure.  Murphy et al. (2009) conclude that the radio deficit regions are approximately coincident with the galaxies' leading edges, where the ICM is actively exerting force on the ISM.  Since the galaxy's tail, which should be roughly aligned with the ICM wind direction, is on the side of the galaxy opposite the radio deficit region, we feel confident calling the side of the galaxy with the radio deficit region ``upwind" and the side with the tail ``downwind" of the major axis.  

Gravitational interactions cannot account for the undisturbed stellar disk (see Section \ref{oldstellardisk}) and disturbed gas distribution in NGC 4330.  In any case, there are no extremely close neighbors that would be likely to cause strong gravitational interactions.  The only Virgo galaxy within 25$\arcmin$ ($\sim$100 kpc projected) and 500 km s$^{-1}$ of NGC 4330 is VCC 0706, an early-type dwarf galaxy at a projected distance of 77 kpc and velocity difference of 75 km s$^{-1}$.  Since it is 4.2 magnitudes fainter than NGC 4330 (B$_T$ = 17.3, and NGC 4330 has B$_T$ = 13.1), it is too small to significantly perturb NGC 4330.  The nearest large galaxy (B$_T$ = 12.5), the spiral NGC 4313, has a projected distance of 130 kpc and a line-of-sight velocity difference of 126 km s$^{-1}$.  It could be close enough in space and velocity to be physically associated with NGC 4330, possibly causing tidal interactions, though the old stellar disk of NGC 4313 appears regular and undisturbed.  Since NGC 4330 is in a cluster environment, we also cannot rule out the possibility that it has been affected by fast gravitational encounters with other galaxies which are now located far away.  However, such encounters could not be responsible for the highly disturbed ISM of NGC 4330, since its stellar disk appears undisturbed.

\section{Observations}
\label{observations}
\subsection{Optical Imaging}
\begin{figure*}[t] 
   \centering
   \includegraphics[width=6in]{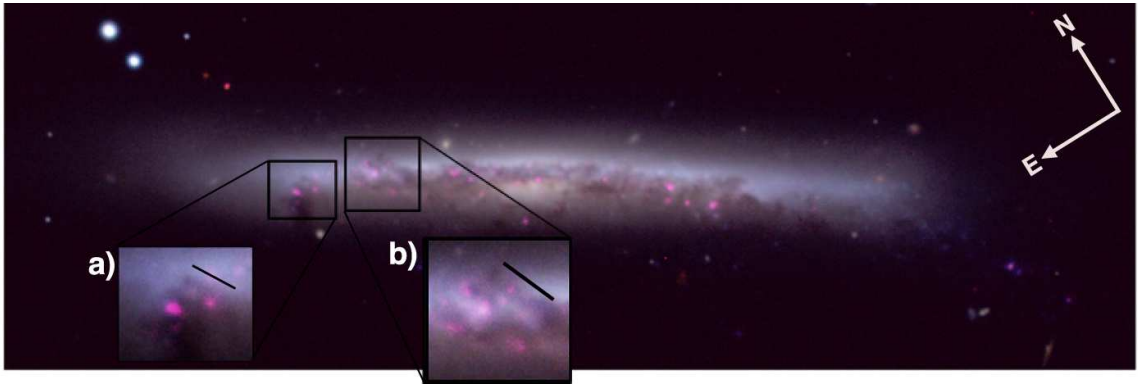} 
   \caption{BVR-H$\alpha$ image of NGC 4330 with insets of interesting dust features.  Each inset has had its contrast adjusted to best highlight the dust features, and the approximate position angle of the linear dust features is indicated with a black line.  In a) and b), we detect dust features which are elongated roughly parallel to the HI tail and perhaps the projected ICM wind direction.  There are also several bright HII regions, which may be in the process of being stripped of obscuring dust.}
   \label{4color}
\end{figure*}

Exposures of NGC 4330 were taken in Harris BVR and narrow-band H$\alpha$ +
[NII] (W16) filters at the 3.5 m WIYN telescope at KPNO on February 3,
2006.  The 4k x 4k OPTIC CCD was used, which has a plate scale of
0$\farcs$14 pixel$^{-1}$, giving a field of view of about 9$\farcm$3, or 45 kpc
at Virgo distances.  The narrowband filter had a bandwidth of 72 \AA \,
 about the central wavelength of 6618 \AA.  Integration time totaled
9 minutes each for R and V, and 15 minutes for B and H$\alpha$ + [NII],
using three exposures per filter.  The seeing in R was 0$\farcs$7, and the
seeing in BV and H$\alpha$ + [NII] was 1$\farcs$1.  Standard reductions
were performed using IRAF\footnote{IRAF is distributed by the National Optical Astronomy Observatory, which is operated by the Association of Universities for Research in Astronomy (AURA) under cooperative agreement with the National Science Foundation.}.  A continuum-free H$\alpha$ +
[NII] image was produced by subtracting the sky-subtracted, scaled R-band
image from the narrow band image.  We also include a 4-color image (Figure \ref{4color}) to illustrate the galaxy-wide dust distribution and how it relates to the older stellar disk and H$\alpha$.  This was constructed by first convolving the R-band to the same seeing as our B, V, and H$\alpha$ images.

\subsection{HI 21 cm Line and 20 cm Continuum Observations}
HI 21 cm line and 20 cm radio continuum observations were taken as part of the VIVA (VLA Imaging
survey of Virgo galaxies in Atomic gas) survey (for futher details see
the VIVA atlas paper, Chung et al. 2009), using the C-short and D-array
configurations.  Observations were taken on August 8, 2005, and on
December 1, 2005, respectively.  The observation time was 8 hours in
C-array and 2 hours in D-array, with a bandwidth of 3.125 MHz centered at a
heliocentric velocity of 1565 km/s.  The velocity resolution of both cubes is 10.4 km/s.  The synthesized beam size FWHM for the C-short array data is 16.5 x 15.9 \arcsec, and the FWHM for the D-array data is 72.2 x 48.2 \arcsec.  Each data set has its own advantages: the C-short array is less sensitive to extended emission, but its superior angular resolution enables more meaningful comparisons between the spatial distribution of HI and the better-resolved maps at other wavelengths.  In order to combine the angular resolution of the C-array and the sensitivity of the D-array, we have constructed a C+D-array cube with a FWHM of 26.4 x 24.0 \arcsec.  In this paper, we use the C-array data to compare the spatial distribution of HI with other wavelengths.  We use the C+D array data for calculating the total HI flux and mass.  The data were calibrated and
continuum subtracted using the NRAO Astronomical Image Processing
System (AIPS).  For further details on radio image processing see Chung et
al. (2009).  

\subsection{GALEX UV Photometry}
We imaged NGC 4330 with GALEX (Martin et al. 2005) with the NUV broadband (1750-2800 \AA) and the FUV broadband (1350-1705 \AA)  filters on May 13, 2007, with exposures reaching a depth of 1.7 ksec in both bands (GI Program Number 79, Cycle 1).  The FWHM is 5.3\arcsec\ for the NUV filter and 4.2\arcsec\ for the FUV filter, and the pixel size is 1.5 arcsec.  Data were processed using the standard GALEX pipeline (Morrissey et al. 2007).  

\section{Results}
\label{results}

\begin{figure}[h] 
   \centering
   \includegraphics[width=3in]{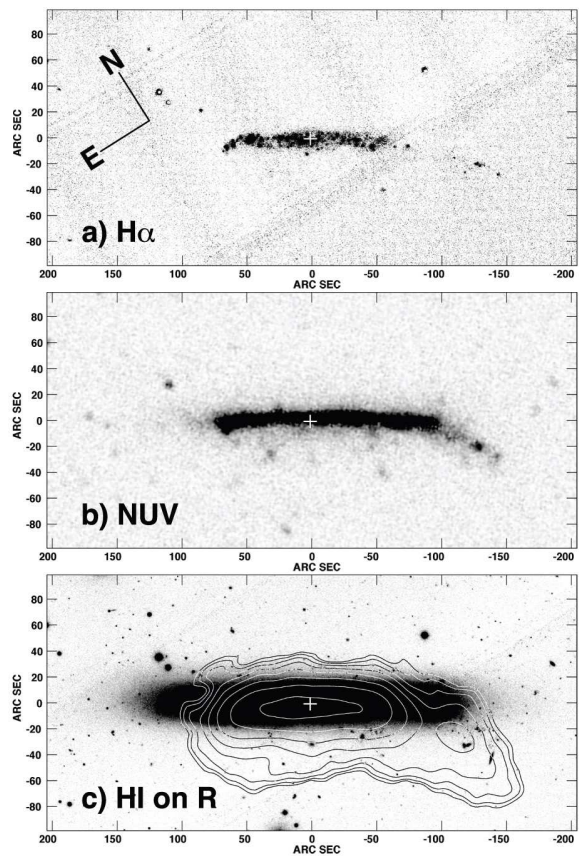} 
   \caption{a) NGC 4330 H$\alpha$ + [NII] greyscale image of NGC 4330- ongoing star formation is largely confined to the inner disk, with a small amount in the tail;  b) NUV greyscale - recent star formation is slightly more radially extended than HI;  c) HI C+D-array contour on R greyscale - the older stellar disk is undisturbed, but HI has been removed from the galaxy's outer disk (contours are 2, 4, 8,... 256 $\times10^{19}$cm$^{-2}$)}
   \label{3panel}
\end{figure}

The undisturbed outer stellar disk, traced in R-band (Figures \ref{3panel}c and \ref{ronhalpha}), indicates that the galaxy has not been significantly affected by gravitational processes, leaving hydrodynamic processes as the likely cause of NGC 4330's truncated star-forming disk and disturbed, asymmetric HI distribution.  Ample evidence indicates ongoing ICM-ISM stripping, which is expected to remove a galaxy's ISM from the outside in and produce a one-sided tail of extraplanar gas while leaving the old stellar disk intact.  Ram pressure may also initiate a burst of star formation along the ICM-ISM interaction boundary and/or in the stripped gas.  Below, we describe the evidence that various components of the galaxy's ISM are experiencing ram pressure stripping. 

\subsection{The Old Stellar Disk}
\label{oldstellardisk}

\begin{figure*}[h] 
	   \centering
	   \includegraphics[width=6.5in]{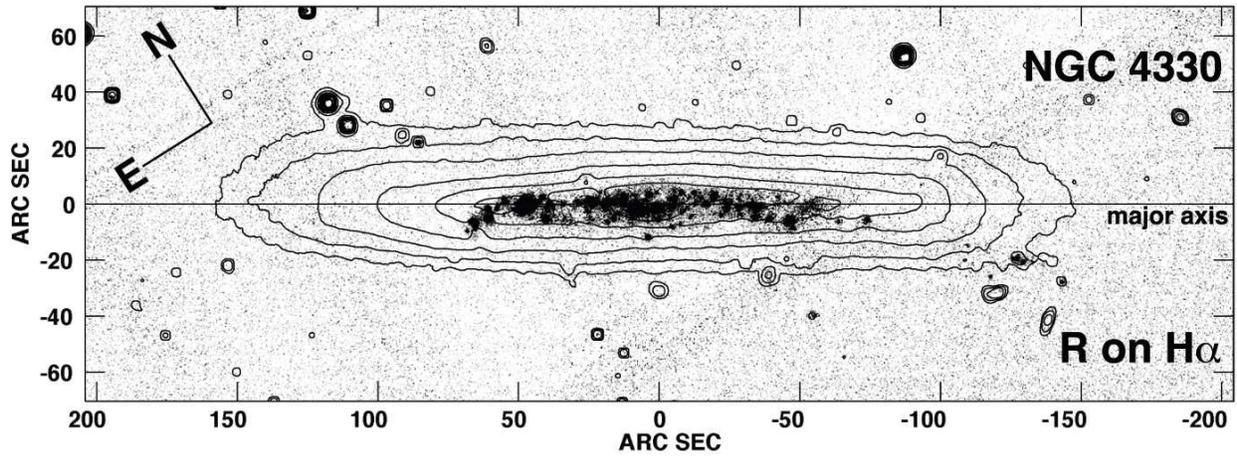} 
	   \caption{NGC 4330 R contours (convolved with a Gaussian kernel to 3$\farcs$ resolution) on H$\alpha$ greyscale.  The outer R-band contours are roughly elliptical and regular (except for point sources).  Each contour level represents a factor of 2 increase in flux from the previous contour, and the outermost contour is at 1.1 R$_{25}$ (25.6 mag arcsec$^{-2}$).}
	   \label{ronhalpha}
	\end{figure*}

The inner disk of the galaxy shows modest asymmetries in R-band light due to asymmetries in star formation and dust extinction, but the outer stellar disk beyond the dust and star formation ($\sim$75$\arcsec$ from the galactic center) is quite symmetric and regular.  The R-band disk remains symmetric out to the deepest extent to which we can detect emission ($\sim$ 11.8 kpc (2.5'), or 110$\%$ of $R_{25}$).  The outer R-band isophotes are roughly concentric (Figure \ref{ronhalpha}), and there is no evidence of any high- or low-luminosity bridges, warps, or other features associated with gravitational interactions.  The rotational period at this radius is $\sim10^9$ years, so if the outer stellar disk had been significantly perturbed in the past few Gyr, the evidence would not yet have been wiped out by the galaxy's differential rotation.  Although there is a low-luminosity tail of young star clusters roughly coincident with the HI tail (see the deep B-band image in Fig. \ref{bdeep}), there is no smooth tail of older stars which, if present, would be a signature of gravitational interaction.

\subsection{Dust Extinction Morphology}
\label{dustmorphology}
\begin{figure*}[htbp] 
   \centering
   \includegraphics[width=6.5in]{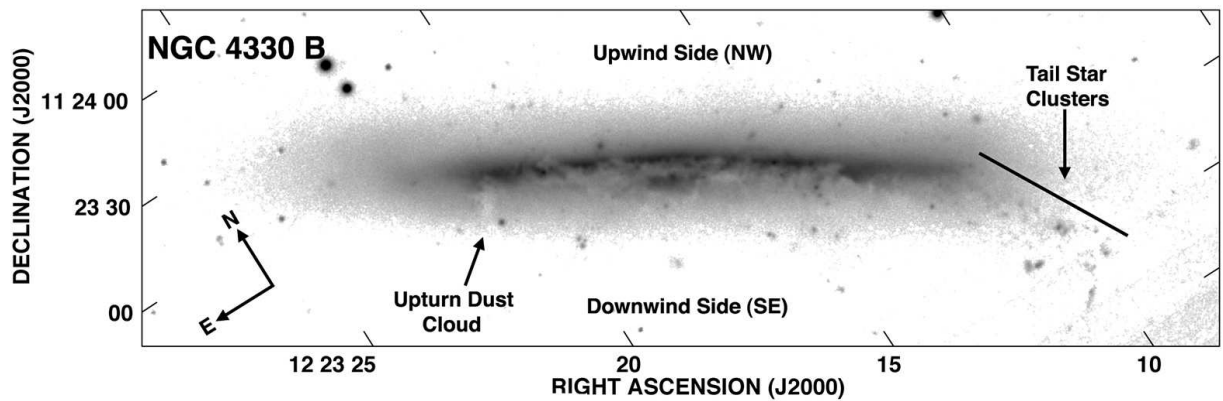} 
   \caption{NGC 4330 deep B-band image.  Note the star clusters in the tail region and the upturn dust cloud.}
   \label{bdeep}
\end{figure*}

\begin{figure*}[htbp] 
   \centering
   \includegraphics[width=6.5in]{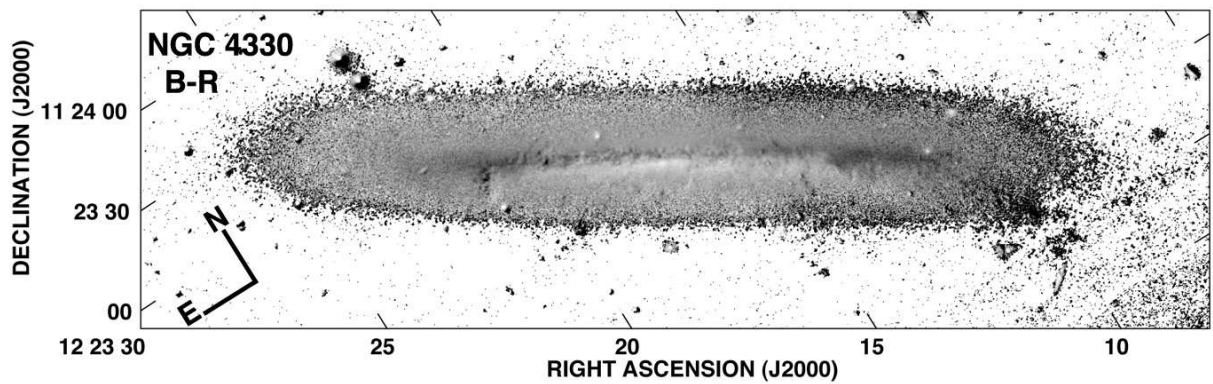} 
   \caption{NGC 4330 B - R image, where darker areas indicate relatively strong B-band emission and light areas indicate relatively strong R.}
   \label{bminusr}
\end{figure*}

We have used the R-band images for analysis of the galaxy's overall dust distribution due to its higher resolution in our data (0\farcs7 seeing in R vs. 1\farcs1 in BV), although we also use B and V bands for our analysis of the morphology of individual features.  
On the southeastern side of the disk, a spatially large ($\sim 10
	\arcsec$ or 800 pc in its longest dimension) and significantly
	obscuring dust plume is elongated almost perpendicular to the
	galaxy's photometric major axis (Figure \ref{bdeep}).  This cloud seems to mark the
	truncation boundary of the dust extinction on the northeastern side
	of the galactic disk.  Beyond it, no extinction is visible in
	the BVR bands, even in the B-R image (Figure \ref{bminusr}) which emphasizes the galaxy's dust distribution while eliminating the gradient caused by the disk's radial decline in luminosity.  We find that at roughly the same radius as the dust plume referred to in Figure 
\ref{bdeep} as the "upturn dust cloud," UV and H$\alpha$ emission are also truncated, curving away from the disk midplane and towards the large dust cloud, in a feature we call the ``upturn" (discussed further in Section \ref{upturnsection}).   
	
	On the northeast (leading) side of the galaxy, there are at least two smaller-scale dust clouds with elongated morphologies (Figure \ref{4color}a, b).  One end of each of these clouds is located upwind (northeast) of the galaxy's continuous dust distribution, and they extend downwind (southwest) to the galaxy's main dust lane.  The cloud in Figure \ref{4color}a is $\sim$6$\arcsec$ (480 pc) long before it becomes indistinguishable from the galaxy's main dust lane, and its maximum width is 1.5$\arcsec$ (120 pc).  The cloud in  Figure \ref{4color}b is $\sim$7$\arcsec$ (560 pc) long by 2.5$\arcsec$ (200 pc).  These may be dense clouds which decoupled from the stripped lower-density ISM and are now being ablated (see Section \ref{windangle}).  
	
	The dust distribution on the trailing (southwest) side of the galaxy at the base of the tail
	also curves away from and out of the disk.  However, it curves much more gradually than the sharp cutoff we observe at the "upturn".  Dust is visible out to about 50$\%$ of $R_{25}$ on the leading (northeast) side, and there is some dust extinction out to approximately 90$\%$ of $R_{25}$ on the opposite, trailing (southwest) side, although it is offset to the southeast of the major axis beyond about 50\% of $R_{25}$.  No extended dust structures are visible northeast of the major axis.

\subsection{HI Distribution}
\label{hidist}
	
\begin{figure}[h*]
\centering
\includegraphics[height=8in]{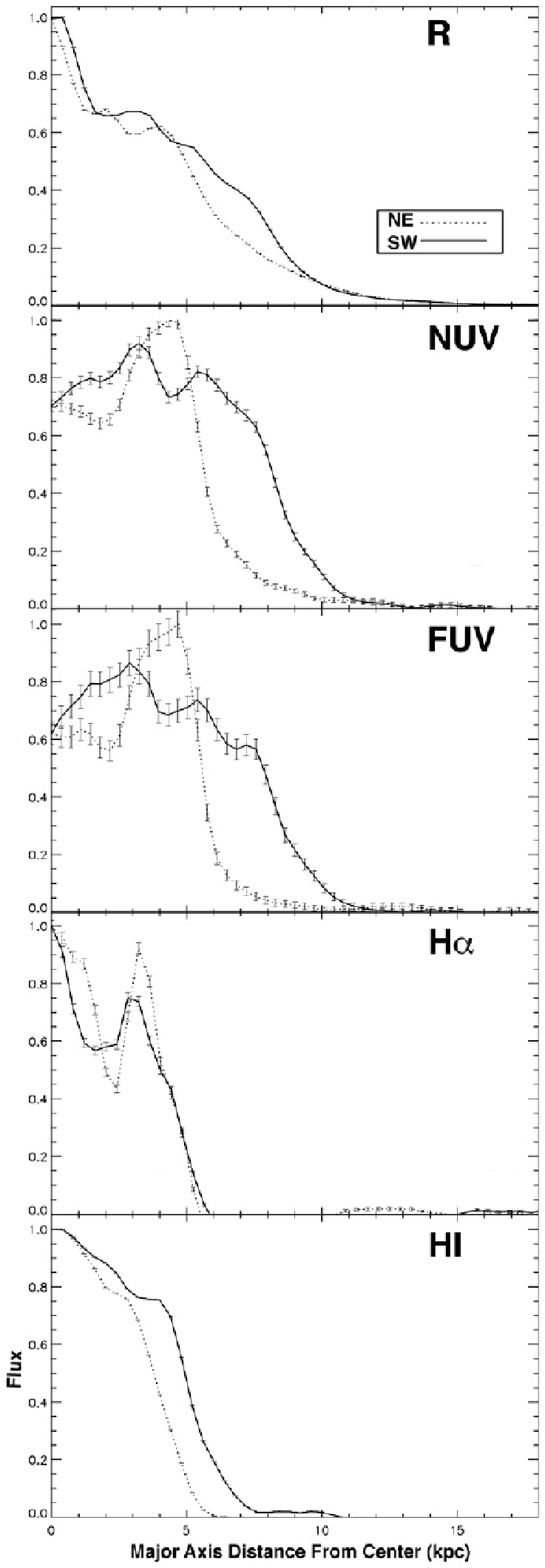} \caption{Normalized flux along NGC 4330's major axis in HI (C-array), R, H$\alpha$, NUV, and FUV.  Images were convolved to 16\arcsec  resolution to match the HI.  The flux is measured in a 15$\arcsec$ strip along the major axis, and binned in 5$\arcsec$ increments, then normalized.  Error bars result from Poisson and sky noise.  Compared to R, H$\alpha$ is strongly radially truncated but symmetric about the galactic center.  HI is slightly less symmetric, and more extended on the tail side.  NUV and FUV are very asymmetric, indicating rapid evolution on 10 - 100 Myr timescales on the tail side of the galaxy.}
 \label{fluxprofiles}
\end{figure}

	The HI distribution in NGC 4330 is radially truncated within the optical disk along the major axis and visibly asymmetric,
	with extraplanar gas and a tail on the southwestern side
	of the optical disk, and little extraplanar HI northeast of the major axis (Figure \ref{3panel}, bottom panel).   The HI distribution is significantly truncated on the leading (northeast) side of the
	optical disk (Figure \ref{upturn}), extending to 55$\%$ of $R_{25}$ along the major axis (Figure  \ref{fluxprofiles}).  On
	the trailing (southwest) side, the HI distribution extends to 95$\%$ of $R_{25}$ along the optical major axis, although the actual tail, which is located southwest of the major axis, extends further.  In Figure \ref{radiodef}, we define the main body and tail of the galaxy, as observed in HI.  

\begin{figure}[h] 
	   \centering
	   \includegraphics[width=3.5in]{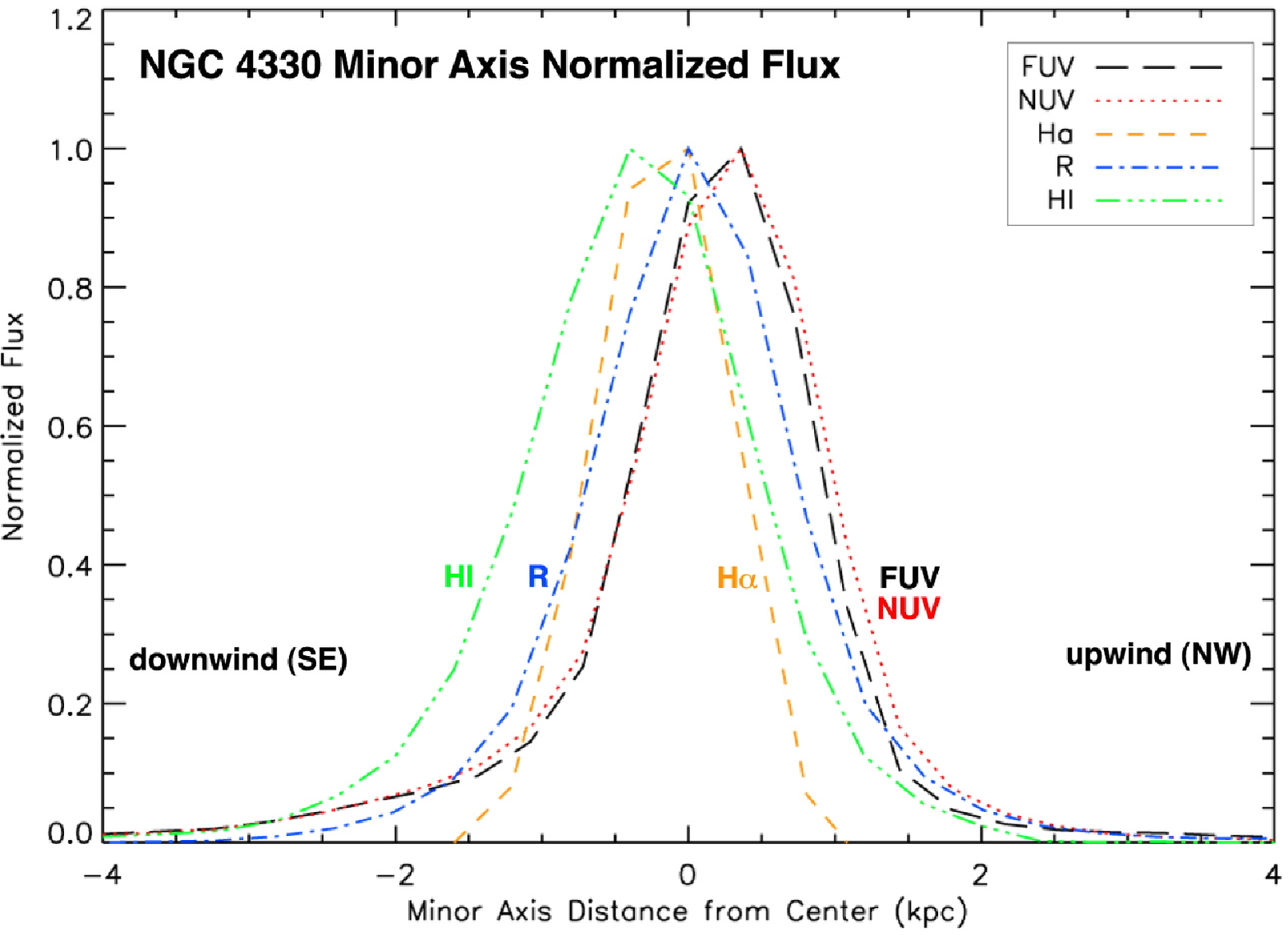} 
	   \caption{Normalized flux along the minor axis of NGC 4330, measured in 5$\arcsec$ bins, where 0 kpc indicates the major axis.  HI emission peaks $\sim$0.5 kpc downwind (SE) of the galactic center, R and H$\alpha$ peak at the center, and FUV and NUV peak $\sim$0.5 upwind (NW).}
	   \label{hicounts}
	\end{figure}

There is a large asymmetry about the major axis in the main body component of HI (region defined in Figure \ref{radiodef}).  The HI distribution peaks downwind of the optical major axis, and there
	is a large excess of emission downwind (southeast) of the galaxy (Figure \ref{hicounts}).  There is little doubt that most of this HI is extraplanar since it is located far outside the disk (Figure \ref{3panel}c), and the HI distribution's asymmetry about the major axis allows us to estimate quantitatively the amount of extraplanar gas.  We find that 22$\%$ (1.4$\times 10^8$M$_{\odot}$) of the gas in the ``main disk" component of NGC 4330 is actually extraplanar (Figure \ref{hicounts}).  10\% of its total HI located in the tail, so a full 32\% of the galaxy's total HI emission is extraplanar.   
	
\subsection{HI Kinematics}
\label{hikinematics}
NGC 4330 is redshifted by 469 km/s with respect to the Virgo cluster's systemic velocity, assuming a systemic velocity of 1100 km s$^{-1}$.  This is not an extreme value compared to other Virgo galaxies, and given that the galaxy has an HI tail pointing roughly away from M87 (Figure \ref{hionrosat}), it is likely that there is a significant plane-of-sky component in the $\sim$northeast direction.  Thus, the galaxy's plane-of-sky motion probably has a component towards the cluster center.  If NGC 4330 has a typical 3D Virgo Cluster velocity of $\sim$1000 - 1500 km s$^{-1}$, then its plane-of-sky component is $\sim$700 - 1300 km s$^{-1}$.  

\begin{figure}[h] 
   \centerline{
   \includegraphics[width=3.5in]{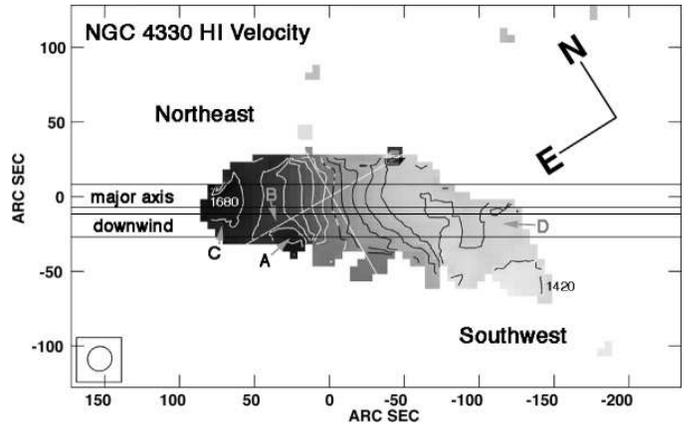} 
   }
   \caption{HI velocity greyscale map of NGC 4330 overlaid with HI velocity contours.  The regions labeled ``major axis" and ``downwind" correspond to the the position-velocity information in Figure \ref{svdmajor}.  The cross mark indicates the kinematic center, with a velocity of 1568 km/s.  The contours are at intervals of 20 km/s.  Key kinematic features A-D are labeled here and in Figure \ref{svdmajor}, bottom panel.}
   \label{svdkey}
\end{figure}

\begin{figure}[h]
   \centerline{
   \includegraphics[width=3.5in]{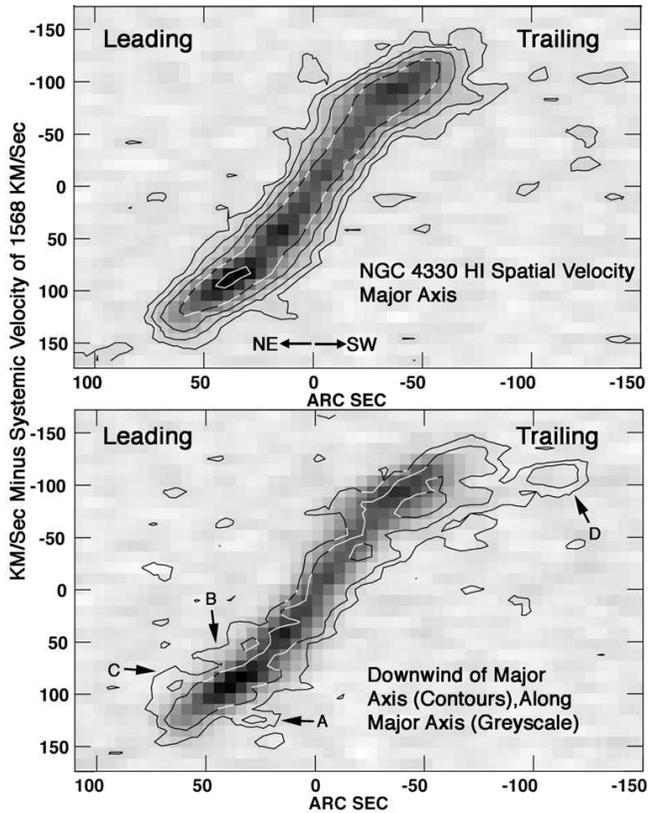} 
   }
   \caption{Top:  Spatial velocity diagram of NGC 4330's major axis, defined as +/- 7.5$\arcsec$ about the galaxy's major axis.  Contours in both panels are 1 mJy/beam x 2, 4, 8, 16, 32.  Bottom:  Spatial velocity diagram of major axis and extraplanar material.
   The greyscale is the same as in the top panel, and the contours are +/- 7.5$\arcsec$ centered 20$\arcsec$ downwind (southwest) of 
   the major axis (see labels on Figure \ref{svdkey} for beam locations).  Key kinematic features A-D referenced in Figure \ref{svdkey} are labeled here.  Contours are same as top.}
\label{svdmajor}
\end{figure}

In Figure \ref{svdkey}, we indicate the regions used in our comparison of the kinematics of planar and extraplanar HI.  The HI along the ``major axis" slice (Figure \ref{svdmajor}, top panel) has relatively symmetric and normal kinematics, which suggests that the kinematics of most of the gas in the disk has not been strongly affected by ram pressure.  The HI along the ``downwind" slice is located in the region of the galaxy where we detect significant excess HI (see Section \ref{hidist}), and its kinematics show signs of disturbance.  

We find that the extraplanar gas on the approaching side of the disk (Figure \ref{svdmajor}, bottom panel) shows a modest ($\sim$ 10 - 20 km/s) systematic offset towards less extreme galactocentric velocities compared to the gas along the major axis.  This differs from some galaxies undergoing active stripping (e.g. NGC 4522, Kenney et al. 2004, Vollmer et al. 2006), where the extraplanar gas has a pronounced overall displacement towards the mean cluster velocity.  Since NGC 4330 is redshifted with respect to the cluster mean, we might expect the stripped gas on the approaching (southwest) side to be accelerated towards us with respect to the galaxy's rotation as it is swept towards us by the ICM along the galaxy's direction of motion.  It is possible that the ICM wind has not affected the stripped gas close to the disk enough to accelerate it to more extreme velocities and the ccgas has lost angular momentum during the stripping process, lowering its galactocentric velocity.  

Figure \ref{svdmajor} (bottom panel) shows that the extraplanar gas on the leading side (northeast of the galactic center) shows little systematic velocity displacement.  However, it does show smaller-scale disordered motions.  We have identified several kinematic features of the extraplanar HI which are robust even to high flux cutoffs, labeled A-D in Figures \ref{svdkey} and \ref{svdmajor}, bottom panel.  In Figure \ref{svdmajor}, bottom panel, feature A indicates that some gas is redshifted away from the cluster velocity at about 30\arcsec\ from the galaxy center, while feature B shows gas blueshifted by $\sim$10 km/s towards the cluster velocity at a very similar radius.  Feature C also shows gas blueshifted towards the cluster velocity.  These disordered motions suggest turbulence in the stripped extraplanar gas on scales of $\sim$20$\arcsec$ (1.6 kpc) and $\sim$25 km/s.    

\begin{figure}[h]
   \centerline{
   \includegraphics[width=3.5in]{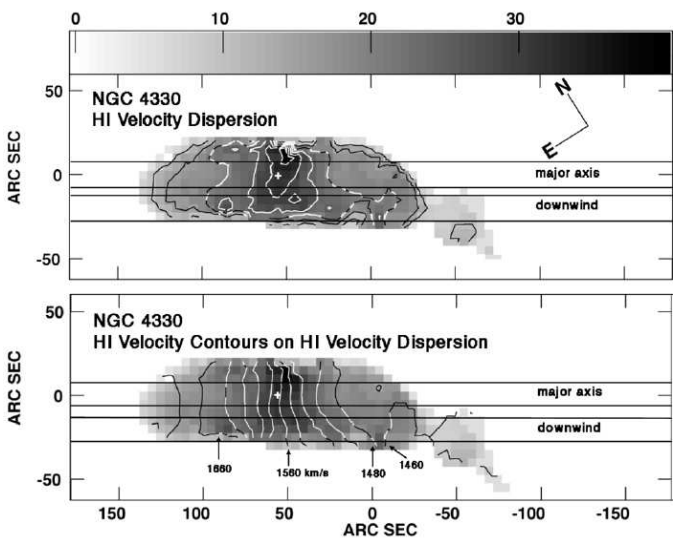}
   }
   \caption{Top:  HI velocity dispersion greyscale and contours for NGC 4330.  Labels refer to same regions as in figure \ref{svdkey}. Contours are 10, 15, 20, 25, 30, 35, 40 km/s.  Bottom:  Same greyscale, HI isovelocity contours.  Contours are 1560 km/s $\pm$20, 40, 60...}
   \label{mom2-on-mom2}
   \end{figure}

A map of the HI velocity dispersion is shown in Figure \ref{mom2-on-mom2}.  The peak of the velocity dispersion is located near the galactic center, as is true for most normal spirals. The velocity dispersion in the tail of NGC 4330 is relatively low, peaking at only $\sim$10 km/s.  Except for the galactic center and the tail, the spectral line width is the same or higher in the extraplanar gas than the major axis gas at a given radius.  However, this effect is modest and may be explained by greater velocity gradients in the extraplanar gas, rather than indicating a true increase in local velocity dispersion.

\subsection{The H$\alpha$ and UV Distributions: Star Formation Tracers}
\label{sftracers}
\begin{figure*}[ht]
\centering
\subfigure{}
\includegraphics[width=2.9in]{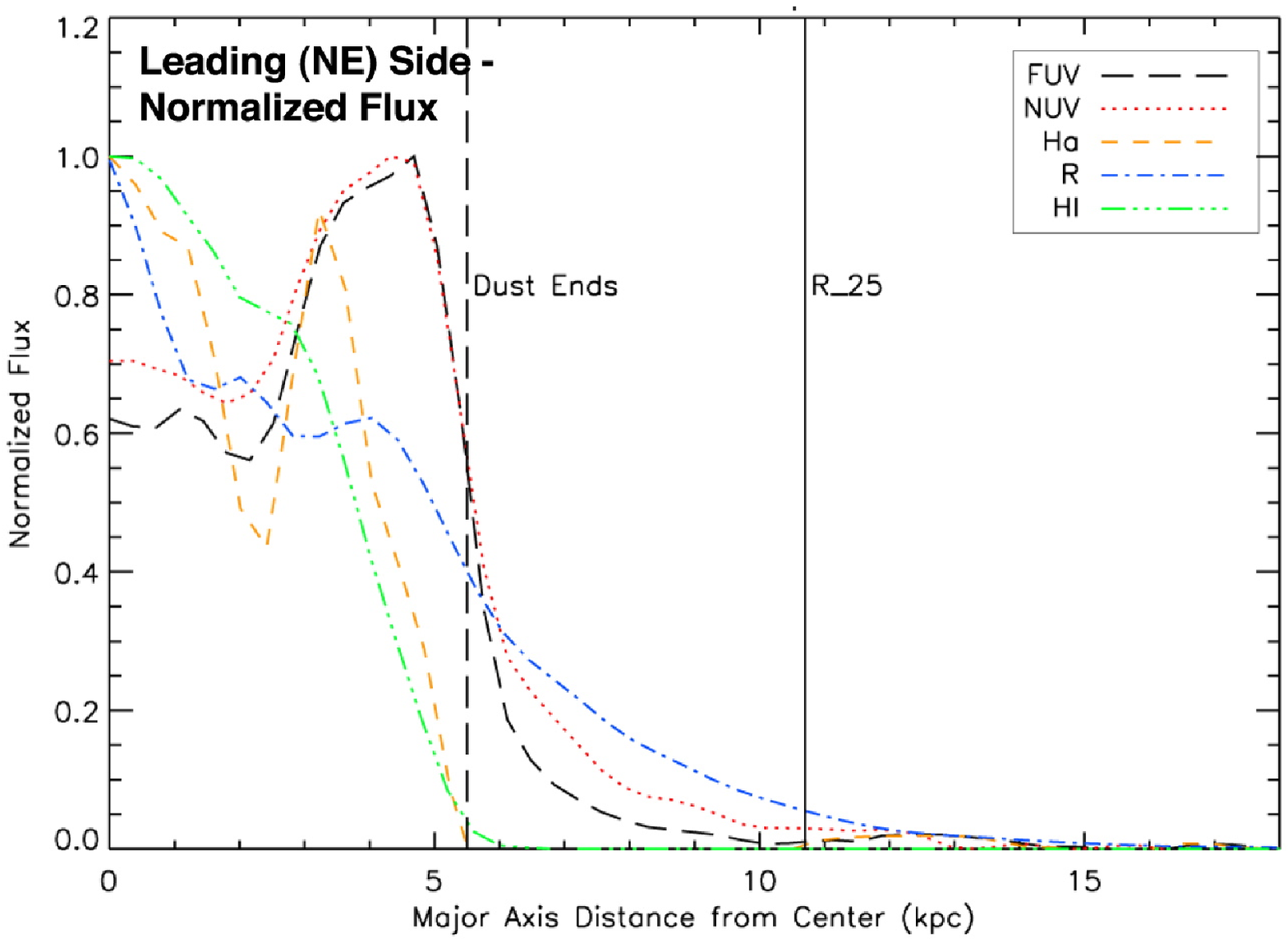} 
\subfigure{}
 \includegraphics[width=2.9in]{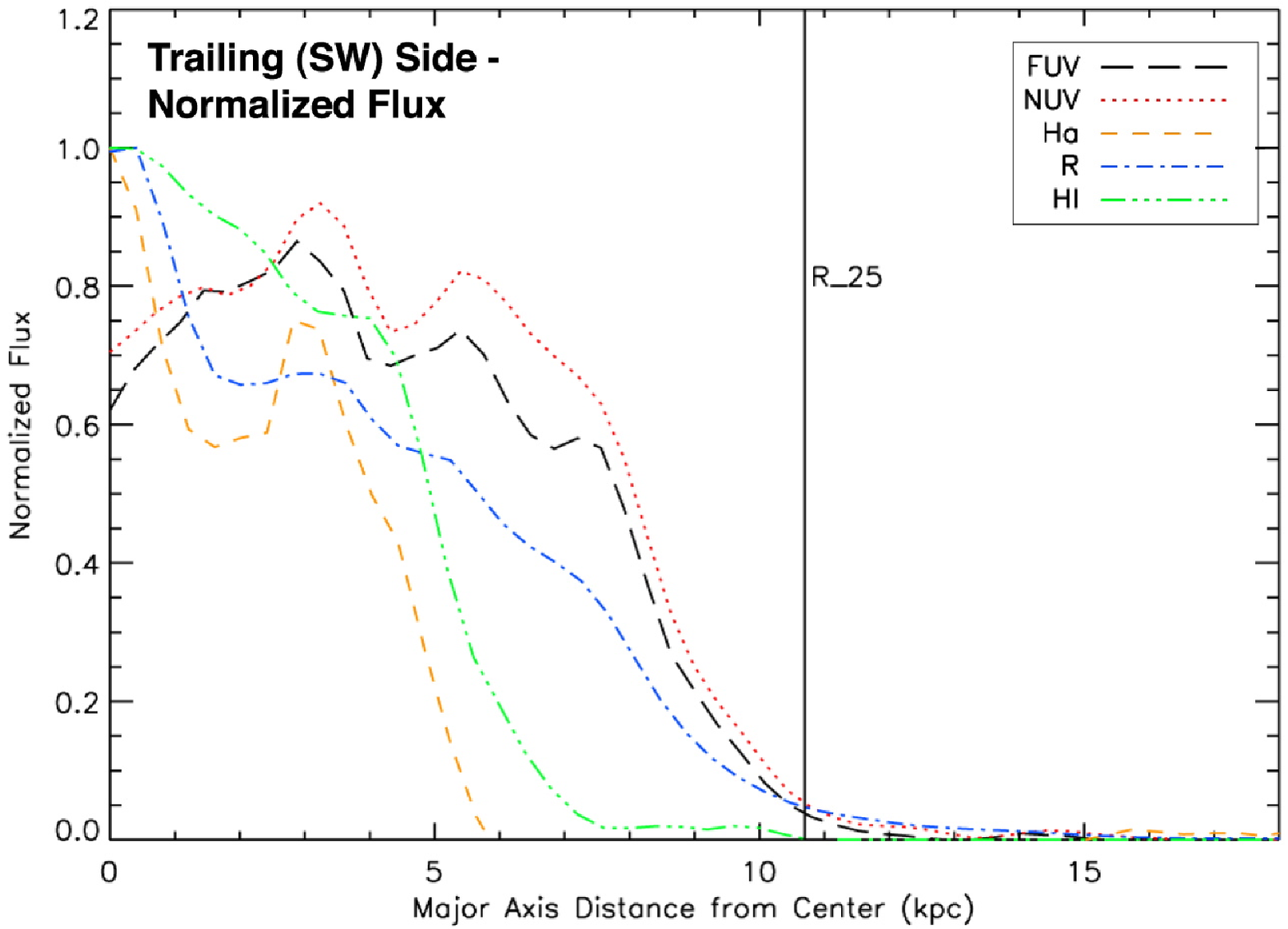}
\caption{Normalized flux along NGC 4330's major axis in HI, R, H$\alpha$, NUV, and FUV.  H$\alpha$ and HI emission on the leading side are truncated at almost the same radius as the dust extinction.}
\label{allfluxprofiles}
\end{figure*}

There is a pronounced asymmetry along the major axis - on the upwind (northeast) side, UV emission is truncated at a significantly smaller radius than on the downwind (southwest) side (Figure \ref{allfluxprofiles}).  There is also a modest excess of flux at H$\alpha$ and UV wavelengths downwind (southwest) of the galaxy, indicating star formation in the stripped material.  The observed asymmetries are strong evidence that the galaxy is being more powerfully affected by the ICM wind on the leading (northeast) side than the trailing (southwest) side.  

The UV in the disk plane extends much further on the trailing side than the leading side  (Figure \ref{fluxprofiles}), with a more contiguous tail structure than is present in H$\alpha$.  On the leading side, NUV and FUV emission peak at 5 kpc and have dropped by 80\% at 6 kpc.  On the trailing side they drop more gradually, with some faint emission extending to or just beyond R$_{25}$.  On both sides of the galaxy, UV emission tapers off more gradually than the H$\alpha$.  Emission at all observed wavelengths other than UV is much more symmetric.  Along the major axis, the H$\alpha$ flux is truncated at the same radius on the leading (northeast) and trailing (southwest) sides of the center, near 50\% of $R_{25}$ (Figure \ref{fluxprofiles}).  It terminates at the same radius on the leading side as the dust extinction and the HI emission (Figure \ref{allfluxprofiles}).  However, on the trailing side, HI extends a bit further along the major axis than H$\alpha$, and the dust extinction does not have as distinct a boundary.  The H$\alpha$ emission is less extended than the UV both radially and in the z-direction.  The fact that the UV distribution is much more asymmetric than the H$\alpha$ and HI along the major axis suggests that the ISM distribution has recently changed.

\begin{figure}[htp] 
\centering
 \includegraphics[width=3.5in]{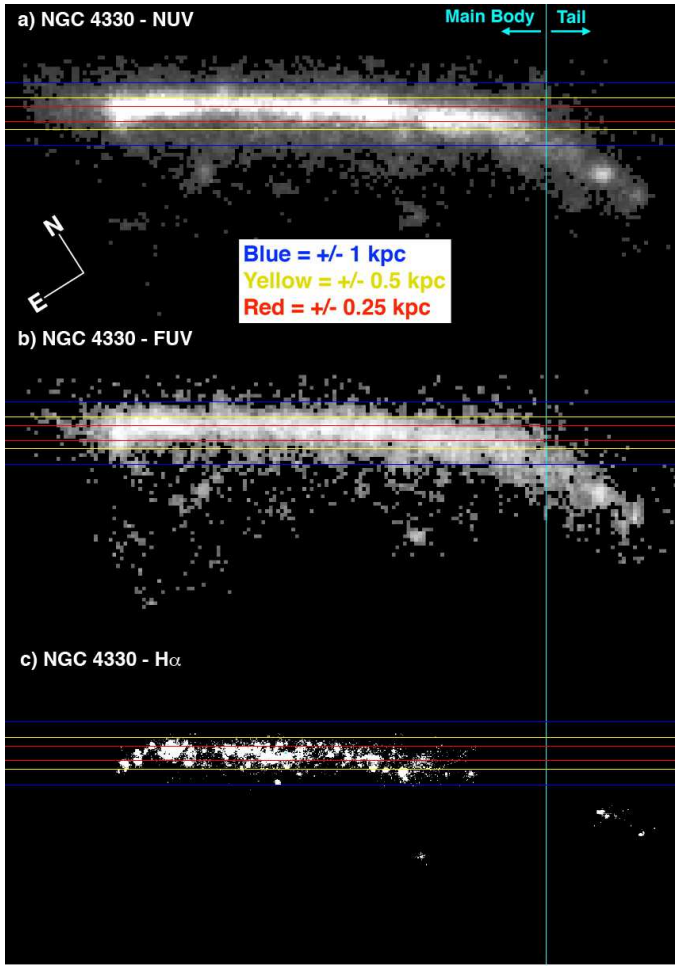}
   \caption{NGC 4330 NUV, FUV, and H$\alpha$, with emission less than 3$\sigma$ above background and emission unassociated with the galaxy masked out.  By excluding the bright central region of the galaxy, we detect excess flux downwind in these wavelengths, a sign of recent and ongoing star formation in the stripped gas.}
   \label{midstripsall}
   \end{figure}

The H$\alpha$ emission is closely confined to the disk plane, but the UV distribution is more extended both upwind and downwind of the disk.  (We discuss extraplanar star formation in discrete regions downwind of the galaxy in Section \ref{xpsfsection}, but in this section we include both these regions and the smooth, continuous emission between 1 - 4 kpc from the major axis.)  In Figure \ref{midstripsall}, we show the UV and H$\alpha$ distributions relative to the major axis.  Of the H$\alpha$ emission outside of the tail, 89\% is located within +/-0.5 kpc of the major axis (between the two red lines in Figure \ref{midstripsall}).  The H$\alpha$ distribution is the least extended of all the wavelengths we examine (Figure \ref{hicounts}).  Its brightness peaks near the galaxy's center rather than upwind, like the UV, or downwind, like the HI.  Excluding the tail, only 0.3\% of the H$\alpha$ is located more than 1 kpc away from the major axis (between the two blue lines in Figure \ref{midstripsall}), compared to 15\% (18\%) of the FUV (NUV).  Emission far from the galaxy due to foreground or background sources has been masked out, and we count only pixels with a surface brightness greater than 3$\sigma$ above background, and count only emission from the main body, not the tail.  

Of the UV emission that is located 0.5 - 1 kpc from the disk plane, there is a significant excess of NUV and FUV flux downwind of the galaxy - 12\% of the galaxy's main body FUV and NUV emission is 0.5 - 1 kpc to the downwind side, versus 3\% (6\%) of the FUV (NUV) upwind.  The flux upwind is likely attributable to the radial extent of the UV within the disk, when viewed at a disk inclination of of 84$\degree$, and is probably not extraplanar.  The excess downwind FUV flux and the high FUV/NUV ratios of the extraplanar starlight indicate recent star formation in the stripped gas.  

As we discuss in Section \ref{xpsfsection}, several discrete UV-bright regions are detected downwind of the galaxy at distances between 1.5 - 9.5 kpc from the major axis, and most of their UV-determined stellar population ages are consistent with populations too old for detectable H$\alpha$ emission.  However, it is also possible, given the ages of some of the regions, that we are detecting young stellar populations traced by UV but not H$\alpha$.  It is well-established (e.g. Thilker et al. 2007) that many spiral galaxies have UV emission without associated H$\alpha$ emission in their outskirts, and this may be due to the same physical conditions that are at work in the extraplanar UV regions in NGC 4330.  The outskirts of spiral galaxies that have UV emission but lack H$\alpha$ are generally lower-density regions than the inner parts of the galaxies, and several possible explanations arise from that fact.  The ionizing photons may leak out of the star-forming regions before creating HII (Oey \& Kennicutt 1997).  Alternatively, the IMF may be undersampled in low-density regions, leading to a shortage of the high-mass O and B stars necessary to create HII regions (e.g. Weidner \& Kroupa 2006), or physical conditions may actually prevent massive stars from forming in low-density regions (Krumholz \& McKee 2008).

\subsection{The Upturn of ISM and Young Stars at the Leading Edge}
\label{upturnsection}
\begin{figure}[h] 
   \centering
   \includegraphics[width=3in]{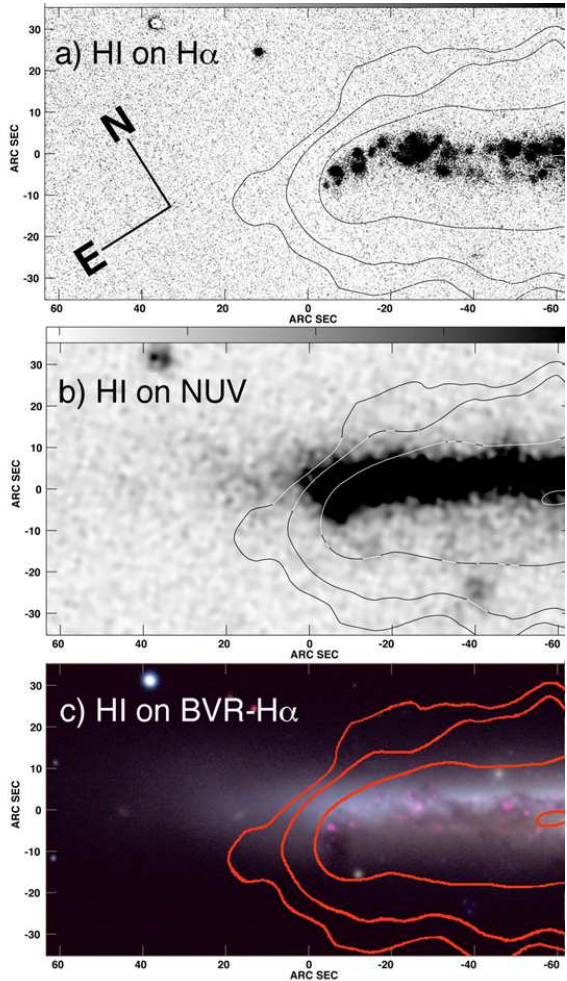} 
   \caption{The northeast (leading) side of NGC 4330, including the ``upturn" region:  a) HI C-array contours on H$\alpha$ - the H$\alpha$ emission is truncated within the HI, and the H$\alpha$ distribution at the galaxy's leading edge curves downwind with a hook-like morphology;  b) HI on NUV - the UV emission extends further radially than the HI and the H$\alpha$;  c) HI on BVR-H$\alpha$ - the R-band stellar disk extends well beyond the emission at other wavelengths, but the dust extinction is confined to much smaller radii.  Note the large, significantly obscuring dust cloud which is located inside the third HI contour.   Contour levels are 2, 4, 16 $\times10^{19}$ cm$^{-2}$.}
   \label{upturn}
\end{figure}

\begin{figure}[htbp] 
   \centering
   \includegraphics[width=3in]{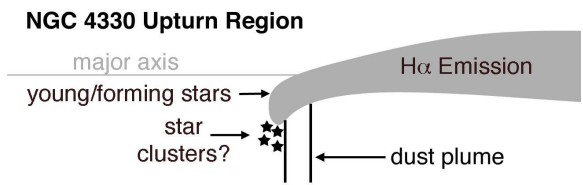} 
   \caption{Schematic of the upturn region in NGC 4330.}
   \label{upturncartoon}
\end{figure}

NGC 4330 has a remarkable feature at its leading edge which we refer to as the ``upturn", shown in Figure \ref{upturn}: at the northeastern H$\alpha$ truncation boundary, the H$\alpha$ emission veers abruptly out of the disk midplane at a $\sim35\degree$ angle.  The H$\alpha$
	upturn has a vertical extent of $\sim 13''$ ( 1 kpc) from
	the optical major axis (Figure \ref{4color}).  The large plume of dust described in Section \ref{dustmorphology} begins at the downwind edge of the H$\alpha$ emission and extends about 0.5 kpc further downwind.  We include a schematic of the upturn structure (Figure \ref{upturncartoon}) to illustrate the relative positions of the different components.  The FUV and NUV distributions gradually taper off in the radial direction beginning at the edge of the H$\alpha$.  A hint of the upturn structure is visible in the HI contours (Figure \ref{upturn}) despite the low HI spatial resolution.
		
The upturn provides a window into the stripping process of the complex, multi-phase ISM.  In Figure \ref{bminusr}, made by dividing the B-band image by the the R-band image, we show that the B-to-R ratio is relatively large on the upwind (northeast) half of the upturn, and relatively small on the downwind (southwest) half.  B is more susceptible to dust extinction than R, and is a better tracer of young stellar populations.  The strongly blue color of the upwind (northeast) half of the upturn structure (indicated in Figure \ref{upturncartoon} as candidate star clusters) strongly suggests the presence of young stars in this area.  Although this region contains some of the bluest B-to-R ratios in the galaxy, the total luminosity of this region is relatively small, and the color could be caused by a relatively small number of young stars.  The redder areas in the B-to-R image are located on the downwind (southwest) side of the upturn, and are coincident with the large optical dust plume (shown in, for instance, Fig. \ref{bdeep}), and the dividing line between high B-to-R and low B-to-R in the upturn may represent the boundary beyond which the obscuring dust has been stripped away from the young star clusters.  

A key as-yet unanswered question about the stripping process is whether the denser parts of the ISM, such as the molecular clouds that could potentially host star formation regions, are stripped from the disk plane at the same time as the less-dense components such as HI.  We detect no H$\alpha$ emission or dust extinction in the disk at radii beyond the upturn, which strongly suggests that at most a very small fraction of the ISM is left behind in this part of the galaxy.  At the same time, the linear dust features described in Section \ref{dustmorphology} are located upwind of the galaxy's main dust truncation boundary, and their morphology indicates that they may be in the process of being ablated by the ICM wind.  The existence of these dust features suggests that although dense clouds can be left behind after the less-dense phases of the ISM are stripped, the remaining clouds are short-lived, which could be why we do not detect any in the disk beyond the upturn.

We think that the H$\alpha$ upturn contains stars which are forming in gas that has been pushed somewhat downwind of the disk plane by the ICM wind.  This would explain the upturn's hook-like shape, as gas further from the galactic center has been displaced further downwind than gas closer to the center.  We interpret the extended dust plume as ISM that has been moved even further downwind, perhaps because it is more susceptible to stripping - it could be more diffuse than the gas that was sufficiently dense to form the stars that now emit H$\alpha$ in the upturn region.

\subsection{ICM Wind Angle}
\label{windangle}

The relationship between the true wind angle $ \alpha$, the angle between the ICM wind and the galaxy's disk plane, and the observed, projected angle $\delta$, the difference between the position angle of the projected ICM wind direction and the position angle of the galaxy's major axis, is given by 
\begin{equation}
\sin  \alpha = \sin \delta \sin i \sin \phi + \cos i \cos \phi
\end{equation}

\noindent  where $\phi$ is the angle between the ICM wind direction and the line of sight and $i$ is the galaxy disk's inclination relative to the line of sight.\footnote{An analogous problem is presented in Kinney et al. (2000), who describe the relationship between the observed and projected orientation of the disk of a Seyfert galaxy and its jet.  The geometry is nicely described in their Figure 2 and Table 4.  We substitute the ICM wind for the Seyfert jet, and define our angles the same way with the exception $ \alpha$, which we define as 90$\degree -$ $\beta$}
 In principle we can observe $i$ and $\delta$, and we can constrain $\phi$ using the galaxy's measured line-of-sight velocity:

  \begin{equation}
\tan \phi =\frac{v_{sky}}{v_{los}}
\end{equation}

  \noindent where $v_{sky}$ is the wind component in the plane of the sky, and $v_{los}$ is the wind component along the line of sight.  In the case of a static ICM, these are equal and opposite to the galaxy's velocity in the plane of the sky and along the line of sight, respectively.  

In the case of an edge-on galaxy with an inclination of $90 \degree$, the second term in Equation 1 vanishes, and the equation reduces to

\begin{equation}
\sin  \alpha = \sin \delta \sin \phi \,.
\end{equation}


We identify three observational constraints on the projected ICM wind angle $\delta$ in NGC 4330: the location of the local radio deficit region, the angle of the HI tail, and the position angle of several smaller-scale, elongated dust features.  The position angle of the large-scale tail will be affected by the stripping history over the past several hundred Myr, while the radio deficit region and the small-scale dust features probably reflect more recent stripping conditions.  When taken into account, NGC 4330's inclination of 84$\degree$ can only potentially decrease $\alpha$ by $\sim$5$\degree$, which we will show is less than the error bars on our measurements.  We therefore treat the galaxy as purely edge-on in the following analysis.   

The local radio continuum deficit region in NGC 4330 (Murphy et al. 2009) is located approximately opposite the HI tail, to the north (upwind) of the galaxy's major axis (Figure \ref{radiodef}).  Murphy et al. (2009) find that  radio deficit regions are located near the leading edges of galaxies experiencing ram pressure.
In the idealized case of a spherically symmetric ISM,
the projected wind angle should be equal to the angle $\theta_{RD}$, the angle between the galaxy's major axis and a line drawn from the galaxy center through the centroid of the radio deficit region.  For a disk galaxy, the ram pressure is more easily able to affect the
low density gas (and weaker magnetic fields) in the halo than the
high density gas (and stronger magnetic fields) in the disk (e.g. Vollmer et al. 2009).
Thus it is easier to create a radio deficit in the halo than the disk, and
as a result the angle between the galaxy center and the centroid of the
radio deficit region is offset toward the minor axis
(the halo-dominated region) from the projected wind angle $\delta$.  Determining this offset quantitatively
would require modelling beyond the scope of this paper.  For this paper, we give rough estimates.

The radio deficit region curves around the disk on the NE (leading) side, extending all the way to the major axis.  The curvature shows that in this region, a radio deficit has been created in material with higher density and stronger magnetic fields than those parts of the galaxy further from the major axis where most of the radio deficit region is located.  Using our definition in the previous paragraph, $\theta_{RD} = $ 60 $\pm 10\degree$.  Since $\delta \le  \theta_{RD}$, as explained above, the projected angle $\delta$ is $<70\degree$.  The curvature of the deficit region means that projected angles down to $\sim$30$\degree$ with respect to the disk plane are possible, so the radio deficit region's morphology suggests a projected wind angle between 30$\degree$ - 70$\degree$.

The position angle of the HI tail is another constraint on the ICM wind angle.  In the case of a static ICM, the centroid of the trail of stripped gas will correspond to the galaxy's orbit through the cluster (e.g., Roediger \& Bruggen 2007).  In models, the relationship between the projected wind angle and the tail angle is influenced by many factors, including the length of the tail being considered, the average density of the tail gas, the evolutionary stage of stripping, whether the model uses hydrodynamics or sticky-particle n-body simulations, and which parts of the tail are used to measure the angle.  In certain cases, the galaxy's tail is fairly well-aligned with the ICM wind direction.  Model snapshots in Roediger \& Bruggen (2006) show that in early-stage stripping ($\sim$200 Myr after the start of pressure), the high-density gas near the trailing side of the disk forms a tail $\sim$20 kpc long whose position angle is within $\sim$15$\degree$ of the ICM wind angle.  The two sides of the galaxy are only asymmetric when the ICM wind is inclined from face-on, so the existence of distinct leading and trailing sides in NGC 4330 is consistent with a wind angle that is inclined from face-on.  No systematic studies of the effect of ICM wind angle on asymmetries in stripped gas exist, but from snapshots in Roediger \& Bruggen (2006) and Roediger et al. (2006), the asymmetry is apparent in galaxies with a wind angle of $60\degree$ (with 90$\degree$ being face-on stripping).  In other papers they treat pure face-on stripping, but a 60$\degree$ angle is the largest non-face-on angle they consider, so we do not know the minimum offset from a face-on wind that produces an identifiably asymmetric HI distribution.  

As stated in Section \ref{thetail}, the HI tail is oriented at a $\sim45 \pm 5\degree$ angle to the disk plane.  It takes time for stripped ISM to move downwind, so we would expect the position angle of the denser tail gas close to the disk to increase over time, and ultimately approach the ICM wind direction (e.g. Section \ref{thetail}).  For a constant ICM wind angle, it seems unlikely that the HI tail near the disk would form a larger angle with the disk plane than the ICM wind does - to first order, the remaining gas disk should shield the recently-stripped material just downwind of it.  Simulations consistently show that the tail's position angle does not exceed the ICM wind angle in the absence of large-scale turbulence or fallback of the stripped material.  The importance of these complicating effects is not yet well-understood, but to first order, we interpret the tail's position angle as an approximate minimum projected ICM wind angle.  As previously stated, we do not know the precise minimum inclination with respect to the ICM wind that produces asymmetric stripping, so we estimate an upper limit on the projected ICM wind angle of $\sim$85$\degree$.  

Assuming that the local gas flow direction within the galaxy is the same as the mean ICM wind direction, the position angle of small, dense clouds being ablated by the ICM wind are likely an indication of the projected wind direction.  The small, elongated clouds in Figures \ref{4color}a and \ref{4color}b, described in Section \ref{dustmorphology}, have position angles of 35$\degree \pm 15 \degree$.  There could be large systematic uncertainties in this method of determining the projected wind angle, since the local gas flow within the galaxy may differ from the mean ICM wind direction, but the position angles of these dust clouds are consistent with the wind angle determined by the two other methods.      


In summary,  the morphology of the radio deficit region indicates a projected wind angle of $\delta$ = 30$\degree$ - 70$\degree$, the HI tail is consistent with a  projected wind angle of 40$\degree$ - 85$\degree$, and the small dust features constrain the local flow direction at their location in the disk to $35\degree \pm 15\degree$.  If all three measurements are considered, the projected wind angle is constrained to 40$\degree - 50\degree$.  Given the unknown systematic uncertainties associated with measuring the wind angle from the dust clouds, we prefer to constrain it based on only the HI tail and the radio deficit region, giving a projected wind angle $\delta$ = $40\degree - 70\degree$. 

The true wind angle $ \alpha$ is a function of both $\delta$ and $\phi$, the angle between the true ICM wind direction and our line of sight.  We can constrain $\phi$ using Equation 2 and estimates for the wind velocity components $v_{los}$ and $v_{sky}$.  The galaxy's line-of-sight velocity with respect to the Virgo Cluster's systematic velocity is +469 km s$^{-1}$, so assuming a static ICM, the ICM wind velocity along the line of sight is $v_{los}$ = -469 km s$^{-1}$ (towards us).  Since the tail is elongated in the plane of the sky, the ICM wind also clearly has a component in that direction, although the magnitude of the plane-of-sky velocity $v_{sky}$ is hard to determine.  Based on the morphology of the tail, $v_{sky} > v_{los}$, so we adopt 469 km s$^{-1}$ as the minimum reasonable value.  No Virgo galaxies have known total orbital velocities $>$ 2000 km s$^{-1}$, so we adopt this as a maximum.  Using Equation 3, we find that $-75\degree < \phi < -45\degree$, where the negative sign indicates that the wind angle is projected in the far hemisphere of the galaxy (due to the negative line-of-sight velocity).  

 We can now constrain $\alpha$ using Equation 1 and our derived estimates of $\delta$ and $\phi$.  Since  $\sin \phi$ will never be $>$1, $ \alpha \le \delta$, meaning that the 3D angle between the disk and the wind will never be greater than the projected angle.  Thus, $ \alpha \le 70\degree$.  A lower limit on $\alpha$ can be calculated using the maximum $\phi$ and minimum $\delta$ values, with the result that 30$\degree\, < |\alpha| < 70 \degree$\footnote{The sign of $\alpha$ is negative in this case because, consistent with the negative value of $\phi$, the wind first encounters the galaxy in the far hemisphere.}.     

\subsection{The Gas-Stripped Outer Disk: Estimation of Stripping Rate from UV-Optical Colors} 
\label{strippingrate}
\begin{figure}
\centering
\includegraphics[width=3in]{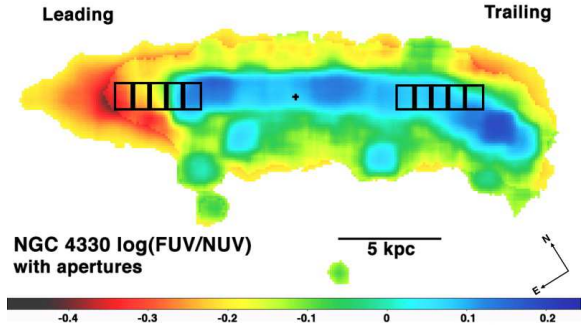}
\caption{log(FUV/NUV) image of NGC 4330, where the input images have been convolved to 9" resolution to minimize noise.  Identifiable background sources have been masked.  Boxes indicate the apertures used to measure flux from to determine the relative UV colors on the leading (NE) and trailing sides (SW).  The apertures are located in the stripped outer disk, beyond most of the galaxy's dust and ISM.  The stellar populations on the leading side have a lower log(FUV/NUV) value, indicating that the stellar populations there are older than the ones on the trailing side.}
\label{fuv-ovr-nuv}
\end{figure}

\begin{figure*}[htbp] 
   \centering
   \includegraphics[width=6.5in]{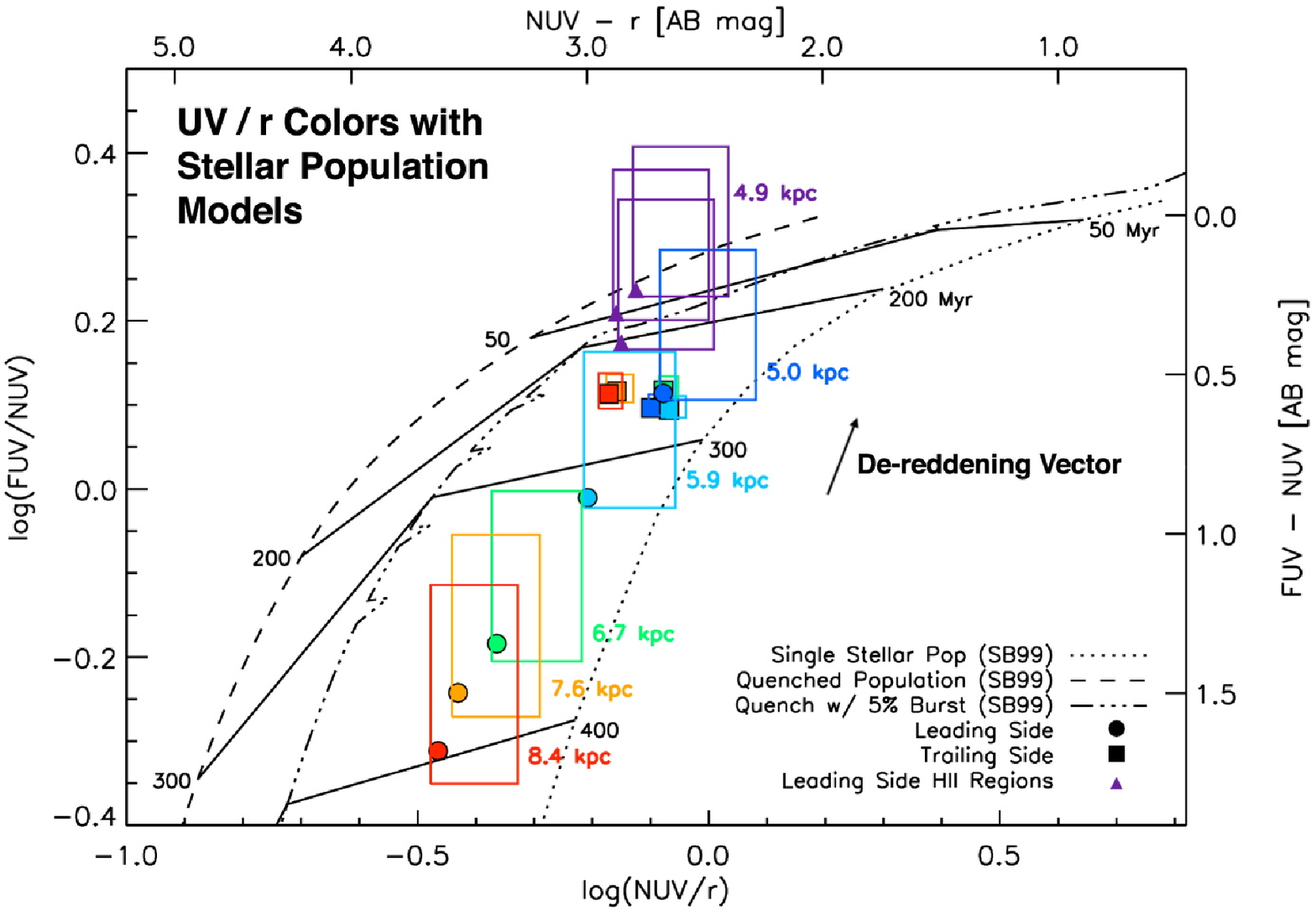} 
   \caption{FUV-NUV vs. NUV-r plot of the outer disk of NGC 4330.  The dashed lines indicate different stellar populations models, from left to right: a quenched model in which constant star formation is abruptly truncated at the specified time, a similar quenched model that experienced a 5\% starburst at the time of truncation, and a single stellar population model.   The solid black lines labeled 50 - 400 Myr indicate a constant age or quenching time across the models.  The points show the data before our correction algorithm has been applied, and the boxes show the range of parameter space which each point may occupy after our correction algorithm has been applied.  The reddening vector has been calculated for A$_{FUV}$=1.  There is a pronounced gradient in stellar population quenching times in the gas-stripped outer disk on the leading side (circles), where the approximate stellar population quenching time increases by $\sim$300 Myr from 5 - 8 kpc in radius, but there is no gradient on the trailing side (squares).}

   \label{sspgrid}
\end{figure*}

The leading side of NGC 4330 has a gas-stripped outer disk between 5 - 8 kpc from the galactic center where H$\alpha$ and HI emission and dust extinction are faint or undetected, but where we can still detect both FUV and NUV emission.  H$\alpha$ emission is a tracer of very recent and ongoing (within $\sim 3 \times 10^6$ years) star formation, and UV light traces young (NUV traces $\sim10^8$ years old or less, FUV $\sim10^7$ years or less) stellar populations (e.g. Bruzual \& Charlot 2005), so the 5 - 8 kpc annular region which lacks H$\alpha$ emission but emits strongly in the UV has had star formation quenched sometime within the past $\sim$0.5 Gyr.\footnote{As discussed in Section \ref{sftracers}, UV emission without accompanying H$\alpha$ is sometimes detected in the outer disks of spiral galaxies, and does not definitely rule out ongoing star formation.  However, the area of NGC 4330 near the H$\alpha$ truncation radius is well within the disk, where we would not expect to find the low-density star-forming regions necessary for ongoing star formation that lacks H$\alpha$ emission.}  Assuming that star formation was occurring at the time of stripping, using the FUV/NUV/optical flux ratios to measure how long it has been since the youngest stars formed in a stripped region will provide an estimate of how much time has elapsed since the area was stripped and star formation was extinguished.  Although UV is very prone to dust extinction, optical images of NGC 4330 show little or no dust extinction on the leading side of the disk beyond the stripping boundary at 5 kpc.  The trailing side shows some dust extinction along the major axis, which can be expected to redden the observed colors.  By comparing the observed UV colors with stellar population models, we can potentially learn how long it has taken to strip the leading edge of the galaxy to its current radius and if the leading and trailing sides of the galaxy have different stripping timescales.

We measured log(FUV/NUV) and log(NUV/r) in 15$\arcsec$ high by 10$\arcsec$ wide apertures along the galaxy's major axis from 5 - 8 kpc (see Figure \ref{fuv-ovr-nuv} for aperture locations).  We ran several different Starburst99 (Leitherer et al. 1999) models in order to determine which type of stellar population best matches the UV-optical colors we observe.  The single stellar population model simulates a stellar population that formed all at once in a burst, and is described by the time since that burst.  The ``quench" model reflects a constant star formation rate, followed by an abrupt truncation of star formation activity, and is described by the time since star formation stopped.  The ``quench with burst" model is similar to the quench model, except that it has a starburst of a given strength at the time of truncation.   In Figure \ref{sspgrid}, we indicate a model in which 5\% of the region's total stellar mass was formed in a burst.  

In the gas-stripped outer disk, there is a large gradient in the UV-optical colors on the leading side, but virtually no gradient on the trailing side (Figure \ref{sspgrid}).  At 5 kpc the colors are very similar on the two sides of the galaxy, with fairly ``blue" colors indicating significant recent star formation.  However, the colors ``redden" significantly between 5 - 8 kpc on the leading side, by $\sim$0.8 AB magnitudes in FUV-NUV and $\sim$0.9 in NUV-r.  This color gradient indicates that star formation has been gradually quenched from the outside of the galaxy inwards over the course of a few hundred Myr on the leading side, and the lack of a similar gradient on the trailing side indicates that star formation was all quenched recently at about the same time.  This difference in star formation histories reflects different stripping histories on the two sides of the galaxy.  Given the similar colors, the trailing side may have been stripped around the time when the gas at 5 kpc on the leading side was stripped.  

Several effects complicate the interpretation of the observed colors, including dust extinction, contamination from the outer disk due to the composite stellar population along a given line of sight, asymmetrical quenching times, and rotation.  These are discussed in detail in the Appendix.  Here we summarize the main results.  

After correcting for the effects of the composite stellar population along each line of sight, we find that our data remain inconsistent with a single stellar population on the leading side (as expected for a disk galaxy), but consistent with star formation that has been gradually quenched from the outside in over the course of 200 - 400 Myr.  The UV/optical color gradient on the leading side is roughly parallel to the model age tracks, strongly suggesting that there is a stellar population age gradient from 5 - 8 kpc.  We are confident that a gradient in quenching times exists in the stripped region on the leading side.  If the gradient were simply a result of the viewing geometry we have attempted to correct for, we would see a similar gradient on the trailing side, but points measured at the same radii on the trailing side all have about the same quenching time (Figure \ref{sspgrid}).  Since there is some dust extinction on the trailing side, we calculate a modest reddening vector of A$_{FUV}$=1, which corresponds to E(B-V)=0.035, using the wavelength-dependent dust extinction formulae of Calzetti et al. (2001).  This gives a quenching time of $\sim$50 Myr ago for the trailing-side regions.  

Just beyond the H$\alpha$ boundary on the leading side, at 5.0 kpc from the galactic center, the corrected colors are consistent with a quenched stellar population with a quenching time of $<$50 Myr.  The modeled burst fraction and quenching time increase with increasing radius.  The maximum quenching time, reached at 8.4 kpc, is 200 - 400 Myr ago.  Even after correction, the burst fractions associated with the outer disk points are much larger than seems realistic, since $\sim$20\% - 30\% of the current population would have been created in the burst.  We suspect that uncertainties in stellar population modeling are at work.    

The recent work of Conroy et al. (2009, 2010 a and b) explores sources of uncertainty in stellar population modeling such as the slope of the IMF, poorly-understood phases of stellar evolution, the effects of metallicity distributions within a population, and dust distribution on large and small scales.  Changes in the detailed treatment of dust extinction or the IMF slope can each cause differences of $>$0.1 magnitude in modeled (NUV - r) colors (Conroy et al. 2010a).  Although Conroy et al. do not analyze the Starburst99 model specifically, the uncertainties they mention apply to all stellar population synthesis models.  We therefore use stellar population synthesis models with an appreciation of the inherent uncertainties.   
 
\subsection{The Tail}
\label{thetail}
\begin{figure}[h] 
	   \centering
	   \includegraphics[width=2.5in]{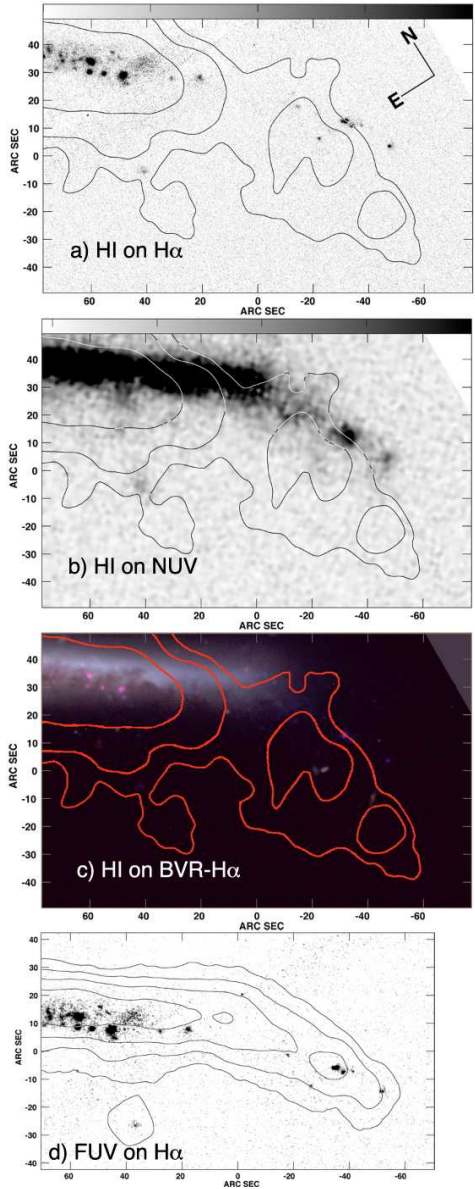} 
	   \caption{NGC 4330 tail and southwest outer disk.  a) HI C-array (highest resolution) on H$\alpha$ - the tail in H$\alpha$ is sparse, consisting of a few discrete regions along the northern border of the HI emission.  Contour levels are 2, 4, 16 $\times10^{19}$cm$^{-2}$; b) HI on NUV - the NUV emission peaks are coincident with those in H$\alpha$, but there is also fainter, continuous emission connecting them with the main body of the galaxy; c) HI on BVR-H$\alpha$ - the dust extinction in the disk decreases significantly at the same radius as the H$\alpha$; d) FUV on H$\alpha$ - the H$\alpha$ peaks are located roughly along the centroid of the FUV tail, suggesting that the location of star formation has not changed in the last $10^7 - 10^8$ years.  }
	   \label{tail}
	\end{figure}

\begin{figure}[h] 
   \centering
   \includegraphics[width=3in]{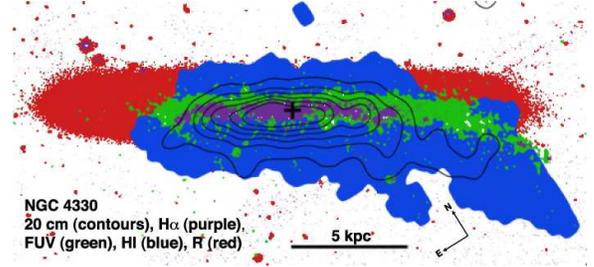} 
   \caption{NGC 4330 schematic showing the distributions of R (red), FUV (green), H$\alpha$ (purple), HI (blue), and 20 cm radio continuum (contours).  The radio continuum and HI tails are displaced downwind (SE) of the UV and H$\alpha$ tails, indicating that the ISM in this area has been pushed downwind over time.}
   \label{everythingoverlay}
\end{figure}

NGC 4330 has a long HI tail extending southwest beyond the optical disk which contains about 10\% (0.4$\times10^8 M_\odot$) of the galaxy's total HI.   The HI tail is 13 kpc long and 5 kpc wide (Chung et al. 2007), and is oriented at a $\sim$45$\degree$ angle to the major axis\footnote{We measure through the  middle of the highest contour of tail emission and assume that the vertex of the angle is located where the tail begins to bend away from the disk (x=0$\arcsec$ and y=30$\arcsec$ in all panels of Figure \ref{tail})}.  The tail structure is particularly interesting because there is a spatial offset between the HI and UV/H$\alpha$ tails (Figure \ref{everythingoverlay}), suggesting recent evolution, and this is discussed below.
	
There is some ongoing star formation in the tail, but it is sparse, consisting of several
	small HII regions extending southwest from the main,
	continuous H$\alpha$ disk.  The tail accounts for about 2\% of the galaxy's H$\alpha$ flux, much less than the 10\% of the total HI flux coming from that region.  There is a large gap in between the
	emission at the base of the tail region and HII regions at the tip of the tail (Figure \ref{tail}), which are located 2.6 kpc from the major axis.  
	The tail's terminal HII region is located just outside $\sim$1.1 $R_{25}$.  No smooth stellar component of the tail is visible in R, although the dust distribution on the tail side of the galaxy curves downwind (southeast) and out of the disk, and star clusters are visible in optical broadband in the tail region.
	
The UV emission in the tail is stronger and much more continuous than the H$\alpha$ (Figure \ref{tail}a, b), with 8\% of the galaxy's FUV and NUV  flux coming from the tail.  The UV tail has no gaps between the disk and tail emission, in contrast to the large gap seen in the H$\alpha$.  The UV tail is oriented at a $\sim$35$\degree$ angle to the major axis, which is shallower than the HI tail (Figure \ref{everythingoverlay}), although they both begin to bend away from the disk at about the same radius. The HI tail is much longer than the UV/H$\alpha$ tail, extending nearly 3 kpc further from the disk.  The sharp drop in UV surface brightness at the downwind edge of the tail in Figure \ref{tail}b suggests that this is a real difference, and not simply a reflection of our sensitivity limits in HI and UV.  

The centroid of the HI tail emission is located downwind (southeast) of the UV tail, suggesting that much of the ISM has moved from the location of the UV tail to the current location of the HI tail.    The C-array (highest-resolution) HI map (Figure \ref{tail}a) shows two main peaks in the tail HI distribution, which are displaced from the peaks in the H$\alpha$ (Figure \ref{tail}b) and UV emission.  In Section \ref{tailevolution}, we estimate that it has taken the HI gas 10 - 300 Myr to move to its current position downwind of the UV-bright stars.  

The lack of widespread H$\alpha$ emission indicates that there has been little star formation in the past $\sim$5 Myr.  The strong FUV emission and ``blue" FUV-NUV colors (see Section 4.5) reveal that, between $\sim$10 and 200 Myr ago, there
was significant star formation activity in the tail.  We can infer that, in
the intervening time, conditions in the tail region became less
hospitable to star formation - either the existing gas was used up by star formation, or
it was swept away during the stripping process.  The presence of significant amounts of HI downwind (southeast) from the stellar tail supports the latter hypothesis.   

\subsection{Extraplanar Star Formation Regions}
\label{xpsfsection}
\subsubsection{UV Properties}
\begin{figure}[h] 
   \centering
   \includegraphics[width=3.5in]{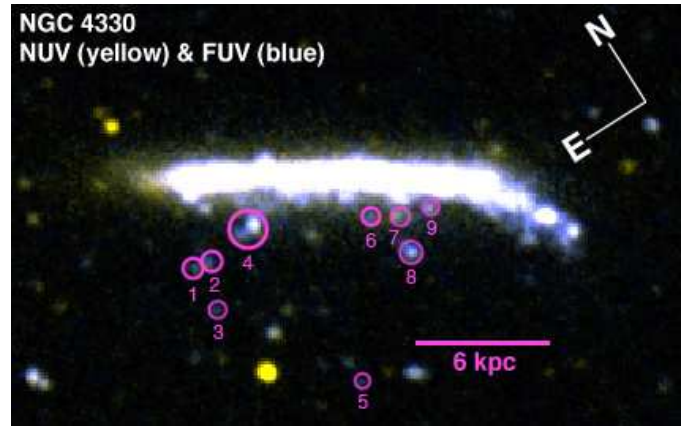} 
   \caption{Composite UV image of NGC 4330 where NUV is yellow and FUV is blue.  We observe 9 extraplanar peaks of UV emission downwind of NGC 4330 which are not coincident with foreground stars or background galaxies in the optical images and presumably originate from recent star formation in gas stripped from the disk.  The circular apertures used to measure the regions' photometry are shown.}
   \label{uvcolor}
\end{figure}

\begin{figure}[h] 
   \centering
   \includegraphics[width=3.5in]{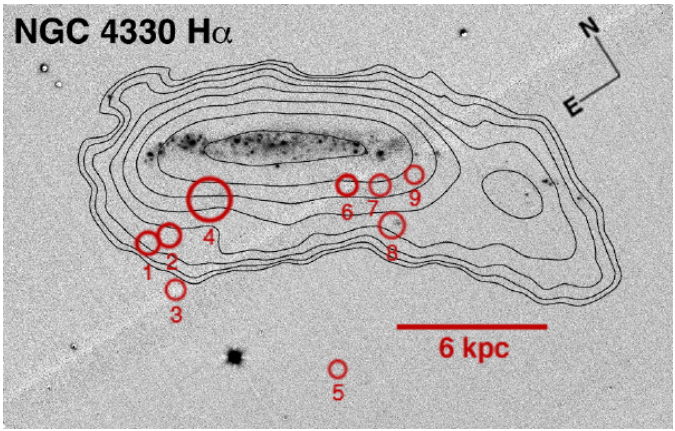} 
   \caption{Continuum-subtracted H$\alpha$ with circles over distinct UV peaks, with HI C+D-array contours.  Although UV emission is detected at significant (3$\sigma$ above background) levels in the H$\alpha$-bright tail, H$\alpha$ is not detected at significant levels in any of the UV-bright regions away from the main body of the galaxy, with the exception of regions 4 and 8.  HI contours are 2, 4, 8,...$\times10^{19}$cm$^{-2}$. }
   \label{hawithaps}
\end{figure}

We detect a striking number of extraplanar UV-bright regions downwind (southeast) of the disk plane, indicating the presence of recent or ongoing star formation in the stripped gas.  There are 9 regions, at distances up to 9 kpc from the disk, which are distinct from the main body and tail of the galaxy and which have surface brightnesses of $\geq$ 3$\sigma$ above background in either FUV or NUV.  We have masked flux from objects whose optical morphologies reveal them to be stars or background galaxies.  Although many of the extraplanar UV-bright regions are faint, they are all on one side of the galaxy.  All but two of the UV-bright extraplanar regions are within the VLA HI contours (Figure \ref{hawithaps}), and one of the remaining regions is close to the outermost detected HI.  We measured the flux of each UV region using a circular aperture as shown in Fig. \ref{uvcolor} and give the results in Table 1.  In total these regions account for 5\% of the galaxy's FUV emission, and 4\% of its NUV.  We detect H$\alpha$ emission downwind of the galaxy from only 2 of the 9 regions, which together comprise only 0.3\% of the galaxy's H$\alpha$ flux.    

Most of the extraplanar UV-emitting regions are in two groups: one is a few kpc downwind (southeast) of the leading edge, extending south beyond the H$\alpha$ upturn (Figure \ref{upturnsection}), and the other is located downwind of the galaxy between the minor axis and the base of the tail.    
While the H$\alpha$ upturn is a continuous curve of HII regions that extends a modest distance of about 0.8 kpc from the major axis, the UV beyond the edge of the upturn extends in a large, arc-shaped complex of regions that reaches $\sim$7 kpc from the main body of the galaxy.  These FUV-bright features may have formed from gas that was originally in the outer disk beyond a radius of $\sim6$ kpc (75\arcsec), a region which is now bright in NUV but faint in FUV.  

\subsubsection{Star Formation Histories of Extraplanar Regions}
The FUV/NUV ratios of these regions can be compared to stellar population models in order to estimate the amount of time that has elapsed since star formation stopped, as well as the stellar masses.  We used Starburst99 stellar population models (Leitherer et al. 1999).  We assume a scenario in which star formation occurred in a short burst in pockets of stripped extraplanar gas, which is best described by a single stellar population (SSP) model.  For the regions closest to the galactic disk (4,6,7, and 9), we have removed the contribution of the smooth disk by subtracting the flux in an aperture of equal size located opposite the major axis from the region itself.  

\begin{figure}[htbp] 
   \centering
   \includegraphics[width=3in]{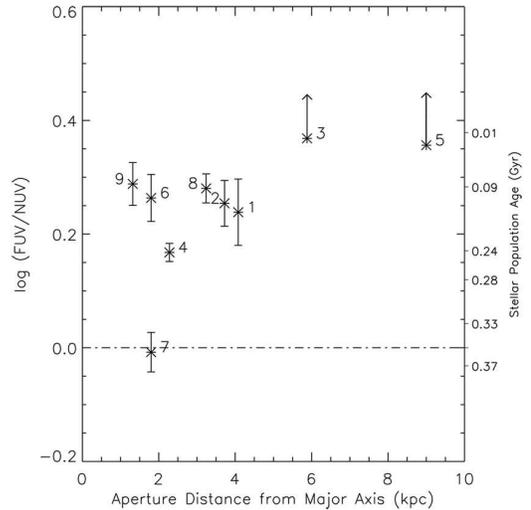} 
   \caption{log(FUV/NUV) for the extraplanar regions in NGC 4330 indicated in Figures \ref{uvcolor} and \ref{hawithaps}, plotted vs. distance to the galaxy's major axis.  Regions 3 and 5 lack significant NUV emission, so their ages are lower limits.  Regions 4 and 8 have associated H$\alpha$ emission, and are probably not well-described by the single-burst stellar population model that provides the ages given on the right-side axis.}
   \label{uvratioplot}
\end{figure}

The FUV/NUV ratios and corresponding ages from the single stellar population model are plotted in Figure \ref{uvratioplot}.  Most of the regions are younger than 100 Myr.  Region 7 is the oldest at 350 Myr, and the second-oldest is Region 4, with an age of $\sim$250 Myr.  These are the two most massive regions, and it makes sense that we only detect old regions with relatively high masses - UV emission becomes relatively faint in older stellar populations, and there are likely other old regions with lower masses that are below our detection threshold.  All of the other extraplanar regions have ages between $\sim$100 Myr at the oldest, and $<$10 Myr at the youngest.  Regions 3 and 5 are detected in FUV but lack significant NUV emission, so assuming an upper limit of 3$\sigma$ above background, we derive upper limits on their ages of $\sim$10 Myr.  The total stellar mass of the regions is 5$\times$10$^{6}$M$_{\odot}$, a tiny fraction of the galaxy's total stellar mass.  

\begin{table*}[t]
   \centering
 \begin{tabular}{@{} c | c | c | c | c | c |c@{}} 
      \multicolumn{7}{c}{Extraplanar UV Region Properties} \\
      \hline 
      Region & FUV Flux & FUV $\sigma$ & NUV Flux & NUV $\sigma$ & H$\alpha$ Flux & Stellar Mass (M$_{\odot}$)\\
      \hline
      1   &  2.80e-17      & 3.4   &    1.58e-17   &  3.8 & $<$4.1e-16   & 1.0e5\\
      2   &  5.43e-17    & 4.4    &    2.97e-17   &  4.7  &    $<$4.1e-16   & 8.2e4\\
      3   &   2.50e-17   & 3.5    &    $<$1.08e-17   &  $<$3.0 &  $<$4.1e-16   & 2.1e3\\ 
      4   &  2.50e-16    & 8.8    &    1.70e-17   &  12.   & 8.58e-16 (4$\sigma$)   & 2.1e6\\
      5   &   1.69e-17   &  3.5   &     $<$7.48e-18  &  $<$3.0 &   $<$4.1e-16   & 5.9e3\\
      6   &  3.64e-17    & 6.6   &    1.99e-17   &  9.1  &  $<$4.1e-16   & 3.9e5\\
      7   &  5.49e-17    &  7.3   &     5.59e-17  &  14.  &   $<$4.1e-16   & 1.6e6\\
      8   &  1.13e-16    & 10.    &     5.88e-17  &  10.  & 2.60e-15 (8$\sigma$)   & 1.6e5 \\
      9   &  6.21e-17    &  11.   &    4.35e-17   &  14.  & $<$4.1e-16   & 1.1e5\\
     
      \hline
\hline
   \end{tabular}\\
   \caption{UV fluxes are in units of erg s$^{-1}$ cm$^{-2}$\AA$^{-1}$.  H$\alpha$ fluxes are in units of erg s$^{-1}$ cm$^{-2}$. }
   \label{sigmatable}
\end{table*}

The oldest of the UV-bright regions, which is one of the closest to the disk plane, is $\sim$350 Myr old, and the quenching time of the outer disk at the galaxy's leading edge is 200 - 400 Myr (Section \ref{strippingrate}).  The removal of gas from the outer disk and the existence of relatively dense extraplanar gas are both consistent with strong stripping beginning around this time.

\subsubsection{H$\alpha$ Properties}
We detect H$\alpha$ emission from regions 4 and 8, to a significance of 4$\sigma$ and 8$\sigma$ above background, assuming point source morphologies.  The detection of H$\alpha$ emission from these two regions may seem somewhat puzzling, since they are not among the youngest regions according to their UV colors.  While H$\alpha$ emission can persist for up to 10 Myr in a single-burst stellar population, our UV-based age estimates of both of these regions are much older, 240 Myr $\pm$ 10 Myr for Region 4 and 90 Myr $\pm$ 15 Myr for Region 8.  

We think the most likely explanation is that the regions with ongoing star formation may also contain dust that was stripped from the galaxy along with the gas that is now forming stars.  The star-forming region emits enough ionizing flux to be detected in H$\alpha$, but dust extinction reddens the UV color by preferentially absorbing FUV rather than NUV photons (as reflected in the wavelength-dependent extinction formulae of Calzetti 2001).  An uncorrected FUV/NUV ratio will yield an age that is older than the region's actual age. 

The SSP models predict an H$\alpha$ flux $<1\times 10^{-17}$ erg s$^{-1}$ cm$^{-2}$ for the brightest region (Region 8) based on the UV-derived masses and ages, which is fainter than our 3$\sigma$ detection limit for a point source.  We do not detect H$\alpha$ in Regions 3 and 5, whose UV colors suggest that they are young enough to have lingering H$\alpha$ emission from a recent episode of star formation, but their predicted fluxes are more than an order of magnitude below our detection limit.   

There are also other scenarios that could also explain UV emission without accompanying H$\alpha$ in low-density or low-metallicity regions (e.g. Lee et al. 2009), so it may not necessarily be a sensitivity issue.  As we discussed in Section \ref{sftracers}, UV emission without associated H$\alpha$ is often detected in the outskirts of spiral galaxies (e.g. Thilker et al. 2005, 2007), which may indicate ongoing star formation with a very low surface density and/or a shortage of the O and B stars which create H$\alpha$ emission.  It is also possible that ram pressure stripping affects the star formation process to such a degree that standard assumptions about the IMF or the formation of HII regions are no longer valid.

\section{Discussion}
\label{discussion}
\subsection{The Evolutionary State of NGC 4330: Comparison With Other Virgo Spirals}
Of the seven Virgo spirals with long HI tails detailed in Chung et al. 2007, NGC 4330 displays the best evidence for strong, active ram pressure stripping.  All of the galaxies in the sample have tails pointing roughly away from the cluster core, which is evidence that they are moving on significantly radial orbits towards the cluster center and are likely making their first pass towards the cluster core.  Since they seem to be falling inwards, they presumably have yet to reach peak ram pressure.  The seven tail galaxies are all located near the cluster's virial radius, but NGC 4330 is the closest, at least in projection, to the cluster core, and may be encountering the densest ICM.  NGC 4330 has the most strongly truncated HI distribution and is the most HI-deficient of the Chung et al. sample galaxies.   

The sample of Crowl et al. (2008) contains several Virgo spirals at more advanced stages of stripping than the Chung sample.  Crowl et al. selected galaxies with star formation, as traced by H$\alpha$ emission, truncated well within the stellar disk.  The galaxies in their sample which are most likely to be analogous to NGC 4330 are the ones whose H$\alpha$ distributions are characterized as ``truncated/normal" (Koopmann \& Kenney 2004b), with no star formation beyond a truncation radius but normal star formation within that radius.  These galaxies also have stellar disks well-fit by elliptical isophotes, with no evidence of asymmetries that might have been caused by gravitational interactions.  Crowl et al. use stellar population analysis to learn the star formation history just beyond the H$\alpha$ truncation radius, indicating when that region was stripped.  The truncated/normal galaxies with young (~50 - 300 Myr) outer disk stellar populations include NGC 4522 (50 - 100 Myr) and NGC 4569 ($\sim$300 Myr).  Both galaxies show significant extraplanar H$\alpha$ and HI, indicating  relatively recent disturbances, and the HI within their disks is highly truncated to $\sim$0.3 R$_{25}$.  

The HI truncation radius of NGC 4330 is further from the nucleus ($\sim$0.6 R$_{25}$) than those of NGC 4522 and NGC 4569, implying that it is at the least advanced stripping stage of the three.  NGC 4330 is less HI deficient than NGC 4522 and NGC 4569.  NGC 4330 has $\sim$6 times less HI than expected for a galaxy of similar size in the field, and NGC 4522 has $\sim$8 times less (Chung et al. 2009).  NGC 4569 is much more HI deficient, with a factor of 30 less HI than expected.  It also has a small HI tail pointing towards M87 (Vollmer et al. 2004), suggesting that it is post-peak ram pressure and therefore a later stage than NGC 4330 and NGC 4522.  

While NGC 4522 has been more strongly stripped of HI, it appears that stripping has been going on for a longer time in NGC 4330.  The large color gradient in the outer disk of NGC 4330 indicates that its stripping has occurred over a relatively long time, 200 - 400 Myr.  In contrast, the relatively blue optical colors of the outer disk in NGC 4522 visible in SDSS images suggest that it was stripped during a recent, rapid, impulsive event.  

Truncated/normal galaxies with old outer disk stellar populations ($\sim$400 - 500 Myr) in the Crowl sample include NGC 4580 and IC 3392.  These galaxies have highly truncated but symmetric HI with regular velocity fields that show no sign of recent disturbance.  There is no evidence of extraplanar HI or star formation, and the galaxies are highly HI deficient.  NGC 4580 has 32 times less HI than a similar undisturbed galaxy, and IC 3392 has 14 times less.  These galaxies probably underwent a similar process to the one currently affecting NGC 4330, but their higher HI deficiencies indicate that more of their HI has been stripped, and their lack of asymmetries or extraplanar material indicates that they experienced peak pressure at least several hundred Myr ago.         

\subsection{The Evolutionary State of NGC 4330: The Orbit within the Virgo Cluster}

The evolutionary state of stripping depends on where the galaxy
is along its orbit through the cluster.
Information on NGC 4330's orbit contributes to a consistent picture
that it is at a pre-peak stage of strong stripping
since it is moving inwards into the central region of the cluster,
probably for the first time.


The projected wind direction of PA = 17$\degree - -13\degree$ (Section \ref{windangle}) measured counterclockwise from north (the number we give in Section \ref{windangle} is the angle between the wind and the major axis) forms an angle of 45$\degree - 75\degree$ with the line between the galaxy
center and M87, where 0$\degree$ is purely radial motion and 90$\degree$ is purely tangential motion.  If the ICM is static, the ICM wind direction
indicates the galaxy's direction of motion on the sky.
Thus, the radial and tangential components of projected motion are similar.
While there is a large tangential component that will carry the galaxy to
the north of M87, there is also a large component which is radially inward.

We can estimate the galaxy's total motion through the cluster.  The galaxy has a line-of-sight velocity of 469 km s$^{-1}$ with respect to the Virgo cluster's mean velocity, and we estimate in Section \ref{windangle} that the galaxy's plane-of-sky velocity is between 500 km s$^{-1}$ and 2,000 km s$^{-1}$.  This gives a total cluster velocity between $\sim$700 - 2,100 km s$^{-1}$.  Based on the stripping history of the outer disk (Section \ref{strippingrate}), gas in the outer disk at radii greater than 8 kpc had been removed by $\sim$200 - 400 Myr ago.  Given the $\sim$55$\degree$ projected angle between the galaxy's orbital path and the line to the cluster center, over a timescale of 200 - 400 Myr, the galaxy has moved 240 - 490 kpc closer to the center.  This means that strong stripping would have started when the galaxy was between 840 kpc - 1.1 Mpc from the cluster center.  In the sample of galaxies with HI tails in Chung et al. (2007), the most distant in projection from the cluster center is 0.9 Mpc.  NGC 4330 could have been at this projected distance or closer when its outer disk was stripped if the stripping timescale is closer to 200 Myr, but the timescale could be 400 Myr as long as its velocity towards the cluster center has stayed under $\sim$ 700 kms$^{-1}$.  It is also possible that ICM substructure led strong stripping to start in NGC 4330 at larger cluster radii than the other tail galaxies - Virgo has significant substructure, and in a dynamic ICM, the ram pressure a galaxy experiences at a given cluster radius can vary by at least an order of magnitude (Tonnesen \& Bryan 2008).  Evidence for the dynamic ICM enhancing stripping has been found in another Virgo galaxy, NGC 4522 (Kenney et al. 2004), which seems to be experiencing ram pressure an order of magnitude stronger than would be predicted by static ICM models for its location in the cluster.  

\subsection{The Evolutionary State of NGC~4330: Comparison With Simulations}
\label{compsims}

A comparison with simulations supports the idea that the ram pressure in NGC~4330
has increased significantly over the last $\sim$300 Myr.  The ISM morphology of NGC 4330 is strikingly similar to the ``medium ram pressure" model scenario of Roediger \& Bruggen (2006) (see their Figures 5 and 7), with p$_{ram}=6.4\times10^{-12}$ erg cm$^{-3}$, t$\sim$200 Myr after the start of ram pressure, and an unprojected wind angle of 30$\degree$ (their Figure 5) or 60$\degree$ (their Figure 7) between the disk plane and the wind direction.  These wind angles are roughly consistent with the $\sim$40$\degree$ - 70$\degree$ range we estimate for NGC 4330 (Section \ref{windangle}).  The model is similar to the stripping history we propose in Section \ref{strippingrate}, which entails a significant increase in pressure 200 - 400 Myr ago.  The leading-edge side of their model also resembles the leading (northeast) side of NGC 4330, with the recently-stripped ISM near the leading side of the disk elongated almost perpendicular to the disk plane, much like the upturn as we detect it in dust extinction.  

It is clear that the stripping process has affected the leading and trailing sides of the galaxy very differently - there is a clear stellar population age gradient on the leading side, but on the trailing side, there is no age gradient and the stellar populations are younger.  The asymmetry observed in this galaxy is also consistent with a significant increase in ram pressure over the last $\sim$200-400 Myr.  Both N-body (Vollmer et al. 2001) and hydrodynamic (Roediger \& Bruggen 2006) models predict that as a galaxy is initially affected by ram pressure, the ISM distribution will become asymmetric because the leading side is affected more strongly than the trailing side.  Over time, asymmetries caused by stripping will even out as each part of the disk rotates through the region of highest ram pressure (Roediger \& Bruggen 2006).  Many details of the observed gas morphology of NGC 4330 are consistent with models with smoothly increasing ICM pressure, in which a galaxy falling into a cluster for the first time starts to experience strong ram pressure over the course of $\sim$200 - 300 Myr (e.g., Roediger \& Bruggen 2006, Vollmer et al. 2001). 

All galaxies orbiting through clusters experience changes in ram pressure strength and direction.
Steady changes are expected if a galaxy orbits through a smooth, static ICM,
although changes may be more sudden in a dynamic, structured ICM.  In NGC~4330, the UV disk asymmetry
indicates a significant increase in ram pressure over the last $\sim$200-400 Myr.  Over 400 Myr, NGC~4330 would likely move radially inward in the Virgo Cluster
from $\sim$0.9 Mpc to $\sim$0.6 Mpc based on typical Virgo velocities.  Over this time interval, the Virgo-like models (with a smooth, static ICM)
of Roediger \& Bruggen (2007) show increases in ram pressure by factors of ~1.3 - 2, to values of P$_{ram}$ = $10^{-11.9} - 10^{-11.5} $erg cm$^{-3}$ for galaxies undergoing pre-peak stripping.  The wind angle can change by $\sim10 \degree$ over 300 Myr for a galaxy which has not yet reached its closest approach to the cluster core, as is likely the case for NGC 4330.  It is not yet clear whether the observed gas and stellar population distributions in NGC 4330 can be explained by steady changes in ram pressure, or whether the sudden changes expected in a dynamic ICM are required.

\subsection{Tail Evolution}
\label{tailevolution}
We observe that the HI tail is offset to the southeast, downwind of the
UV-bright regions. Such offsets between gas and young stars
can provide strong constraints on the recent stripping history of a galaxy.

Spatial offsets between gas and recently-formed stars naturally occur
with gas stripping (e.g., the models of Kronberger et al. 2008).  Whereas the ISM continues to be accelerated by ram pressure,
newly-formed stars decouple from the ISM, and retain the
orbital energy and angular momentum from their birth.  As the rest of the ISM continues to move downwind, it will leave a trail of newly-formed stars in its wake, resulting in a tail of young stars possibly oriented at a different angle from the disk plane than the HI tail.  Small offsets between the newly-formed stars and HI are expected with small or nearly constant ram pressure, but large offsets, like those in NGC 4330, may indicate large and changing ram pressure.  As explained in Section \ref{compsims}, the offset between the UV and HI tails may be consistent with with changes in the ICM wind angle and pressure that occur during a reasonable galaxy orbit.  

We can use stellar population ages of the UV-bright regions to place limits  on the timescale of changes to the ISM distribution in the tail and trailing side of the disk.
We find that the timescale for the HI tail to move from the location of the current UV tail to the current HI position is between 10-300 Myr, depending on whether dense clouds decouple from the rest of the ISM.  If the entire ISM stays together, then the star formation observed in the tail HII regions must have been initiated when HI was located where the HII region is now.
In this case the timescale for the HI tail to move is similar to an HII region lifetime, or $<$10 Myr.  On the other hand, if some dense star-forming clouds decouple from the rest of the ISM (as seen in some other galaxies, e.g. Crowl et al. 2005), but most of the UV-bright stars form within the HI component of the ISM, then the star formation observed in the tail HII regions could have been initiated long after the lower density HI moved downstream. In this case the timescale for the HI tail to move is similar to the time since star formation stopped in the UV tail, or up to 200-300 Myr ago, based on optical-UV colors and
stellar population models.  Finally, if all star formation occurs in (long-lived) dense star-forming clouds which decoupled from the HI (which we consider unlikely), the timescale could be even longer than 300 Myr.

Other effects may be relevant for producing a UV-HI offset, although further models are necessary to gauge the extent to which the tail angle might be affected.  Turbulent motions at the ICM-ISM interface can cause the tail to shift positions, which could contribute to the HI tail's displacement from the UV/H$\alpha$ tail.  Judging from the hydrodynamic simulations of Roediger \& Bruggen (2006, 2007), turbulent motions as the ICM strips away the ISM gas can easily cause a $\sim 15 \degree$ change in the angle between the denser parts of the tail and the major axis over timescales as short as $\sim$10 Myr.  These short timescales are just within the lower limit of the amount of time NGC 4330's tail may have taken to shift positions.  Alternatively, a combination of a non-linear star formation law plus projection effects can produce offsets between the peaks of gas and star formation as observed in 2D maps (Vollmer et al. 2008), but the large-scale, systematic offset we observe in NGC 4330 makes this an unlikely explanation.  Finally, rapid changes in ram pressure in a dynamic, structured ICM may affect the morphology of galaxies undergoing stripping.  There is some evidence that shocks or bulk motions in the Virgo Cluster ICM can increase the ram pressure strength beyond what is expected from a smooth, static model (observations of Kenney et al. 2004; simulations of Tonnesen \& Bryan 2008).  More detailed modelling will be required to test whether the HI-UV offset
implies slow or rapid changes in ram pressure.

\subsection{Extraplanar Star Formation}

Under what circumstances does star formation take place in the material stripped from galaxies, and where does it happen?  Star formation is not sufficiently well-understood to be able to rely on models to predict where in the stripped gas it will take place.  In NGC 4330, for at least some of the extraplanar UV regions, star formation must have occurred after their parent gas clouds left the disk.  For instance, Region 5 (see Figure \ref{uvratioplot} for UV colors and Figure \ref{uvcolor} for location) has a stellar population age of $\sim$10 Myr, and is located $\sim$9 kpc from the galaxy's major axis.  In order for star formation to have started while the cloud was still in the disk, the star-forming cloud would have had to move 9 kpc in 10 Myr, which translates to an average velocity of $\sim$900 km s$^{-1}$.  Region 3 has a similar stellar population age and would require a velocity of $\sim$600 km s$^{-1}$.  While these velocities are comparable to the ICM wind speed, they are probably much larger than the true velocities of the gas clouds, since clouds are not immediately accelerated to the ICM wind speed, and not all gas pushed out of the disk will achieve escape velocity.  The fact that nearly all of the extraplanar star formation regions are associated with HI, which has velocities within 100 km s$^{-1}$ of the disk velocities (see Section \ref{hikinematics} and Figure \ref{svdmajor}), is evidence that these regions are bound to the disk and have much lower velocities.  It is therefore much more likely that at least some of the extraplanar star formation is initiated in extraplanar regions, and not in the disk.

\subsection{New Stellar Components Formed in Ram Pressure Stripped Gas}
In NGC 4330 we observe recent star formation that has occured outside
the thin disk due to ram pressure.  Assuming that these stars formed from gas with roughly the same kinematics as the extraplanar HI, most of these newly formed stars are close enough to the galaxy
that they are likely bound to it, and will form new stellar components
within the galaxy.  On the trailing side, many UV-bright stars in the tail
are displaced from the major axis by as much as 3.5 kpc.
Over a few dynamical times, the stars that are currently UV-bright will spread out azimuthally and vertically
to form a thick disk, with a radial extent of $>12$ kpc (the radius of the stars currently at the tip of the tail).  Stars that form further from the disk, out along the tail or in other stripped gas, and remain bound to the galaxy, will create new halo components. In some cases the distinction between thick disk and halo will be arbitrary.  The H$\alpha$ upturn feature will also create a thick disk component,
although possibly
a much narrower one in radial extent than the one formed by the
broad base of the tail on the trailing side.
An upturn feature may propagate radially inward as stripping advances,
injecting young stars into a thick disk or halo
at progressively smaller radii.

In contrast to thick disks made by mergers,
which are {\it older} than their embedded thin disks
the thick disks made by  ram pressure stripping would be {\it younger}
than their embedded thin disks.
They are also expected to have a narrow range of stellar ages at any location.
The vertical height and age of the stars formed through ram pressure
are expected to vary with radius.
Since galaxies are stripped from the outside in, in general
younger thick disk or halo stars are expected closer to the galaxy center.  The example of NGC~4330 demonstrates that
ram pressure stripping can produce
stellar thick disk and halo components
with distinctive morphologies and age distributions.  Identifying such stellar components hold the potential for revealing the history of past ram pressure interactions in galaxies.

\section{Conclusions}
\label{conclusion}
NGC 4330 provides an outstanding opportunity to study the effects of active ram pressure stripping on a nearby galaxy.  The wind angle, evolutionary stage, and galaxy orientation make it possible to clearly observe distinct behavior at the leading edge (the upturn) and the trailing side (the tail) of the ICM-ISM interaction.  Its stellar disk is undisturbed, with no warps, bridges, strong asymmetries, or other evidence of gravitational interaction.  Ram pressure stripping is the only plausible explanation for its radially truncated, one-sided extraplanar HI distribution and undisturbed old stellar disk.

We conclude by describing the morphological features related to stripping, how we think they were created, and the derived evolutionary timescales for changes in the galaxy over the past $\sim$400 Myr.  Finally we provide a self-consistent description of the timeline of the galaxy's stripping history which takes into account these morphological features and their evolutionary timescales.      

\begin{enumerate}
\item \textbf{Morphological Features Related to Stripping and Derived Stripping Interaction Parameters:}
\begin{itemize}
 \item ISM truncation along the major axis:  The galaxy's ISM has been stripped to well within the undisturbed old stellar disk.  Its dust extinction and HI are truncated at roughly the same radii, $\sim 50\% - 65\%$ of R$_{25}$ on both the leading and trailing sides of the galaxy.  
\item Extraplanar HI:  The galaxy's HI distribution is strongly asymmetric as a result of having been removed from the disk by the ICM - 32\% of the emission is extraplanar on the side we identify as downwind of the major axis.  This includes 10\% of the galaxy's HI in a long tail.  
\item UV major-axis asymmetry:   Recent star formation as traced by UV is quite asymmetric about the galaxy's nucleus - the UV emission strength drops rapidly at 0.5 $R_{25}$ on the upwind side, but gradually tapers off until R$_{25}$ on the downwind side.  The UV emission extends radially further on the upwind side than other tracers of the ISM and star formation - ongoing star formation as traced by H$\alpha$ is truncated at $\sim0.5 R_{25}$, roughly the same radii as the other ISM tracers, and is relatively symmetric about the galaxy's nucleus.  The different extents of H$\alpha$ and UV emission indicate that star formation extended to larger radii when the UV-bright stellar populations formed a few hundred Myr ago, but was much more radially truncated when the H$\alpha$-bright regions formed $<$10 Myr ago.  Based on the UV color gradient in the outer disk on the galaxy's leading side, we estimate that it has taken 200-400 Myr to strip the disk from $>$8 kpc to 5 kpc.  However, there is no color/age gradient on the trailing side, so the effects of stripping have probably been more uniform and recent on the trailing side.
\item The ``upturn":  At the leading edge of the ICM-ISM interaction, the galaxy's H$\alpha$ emission veers abruptly out of the disk midplane, and a large, significantly obscuring dust cloud extends beyond the H$\alpha$ upturn emission at a projected angle almost perpendicular to the disk.  This striking ``upturn" feature is near the HI truncation boundary, and the UV emission extends radially beyond it, suggesting that the ISM in the disk beyond the upturn was removed only recently.  The upturn's H$\alpha$ morphology indicates that most of the ISM in this region has been pushed out of the disk by ram pressure.  The galaxy also displays a strong, localized radio deficit region near the upturn, characteristic of galaxies experiencing strong ram pressure.  
 \item The tail:  NGC 4330 has a long tail on its trailing side which is visible in HI, UV, and H$\alpha$.  The HI tail is offset downwind from the the UV/H$\alpha$ tail, suggesting that much of the ISM has moved from the location of the UV tail to the current location of the HI tail.  There is significant UV emission from the tail but the H$\alpha$ is sparse, indicating that there was significant star formation in the past $\sim$10 - 200 Myr but there has been little in the past $\sim$5 Myr.  We estimate that it has taken the ISM tail $\sim$10 - 300 Myr to move downwind from the position of the UV-bright tail to the current position of the HI tail.  This compares with the galaxy's rotational period of $\sim$400 Myr at a radius of 5 kpc.  
 \item Extraplanar UV-bright regions:  We detect 9 regions downwind of the galaxy but outside the tail that are bright in the UV, but for the most part lack H$\alpha$ emission.  A comparison of their UV/optical colors with stellar population models indicates that most of the regions have ages $<$ 100 Myr, though the oldest is $\sim$350 Myr.  Based on the ages and distances of the most distant regions from the galactic plane and the kinematics of the associated HI, it is likely that star formation in these regions was initiated after the gas left the disk.
 \item The projected ICM wind angle:  We discuss three methods to constrain the projected ICM wind angle
from observations: the morphology of the large scale stripped gas tail,
the orientation of elongated small-scale dust features, and the
shape of the radio deficit region. We give quantitative estimates
for all three techniques in NGC 4330, and constrain its disk-ICM wind
angle to be 30-70$\degree$.

\end{itemize}



\item \textbf{Stripping History:}
The observations are consistent with the following scenario.  Ram pressure increased rapidly $\sim200-400$ Myr ago, stripping gas from the disk on the leading edge side from $>$8 to 5 kpc.  The disk was stripped asymmetrically, and for a while the gas disk was stripped more deeply on the leading side than the trailing side.  After $\sim$1 rotation period, ($\sim$400 Myr) at this higher pressure, the gas disk has become more symmetric, as currently observed.  Within the last $\sim$400 Myr, as the gas disk has gone from highly asymmetric to fairly symmetric, gas on the trailing side has left the disk to form the base of the tail, and the tail has continued to move from the disk.  It is not yet clear whether smooth or abrupt changes in ram pressure are responsible for this galaxy's recent, dramatic evolution.

\end{enumerate}

\acknowledgments
This work was partially supported by the National Science Foundation under grant number 0607643 to Columbia University.  GALEX (Galaxy Evolution Explorer) is a NASA Small Explorer, launched in April 2003. We gratefully acknowledge NASAÕs support for construction, operation, and science analysis for the GALEX mission, developed in cooperation with the Centre National d' Etudes Spatiales of France and the Korean Ministry of Science and Technology.  

\appendix
\section{Appendix: Complicating Effects in UV - Optical Color Interpretation}  

\begin{figure}[htbp] 
   \centering
   \includegraphics[width=2.5in]{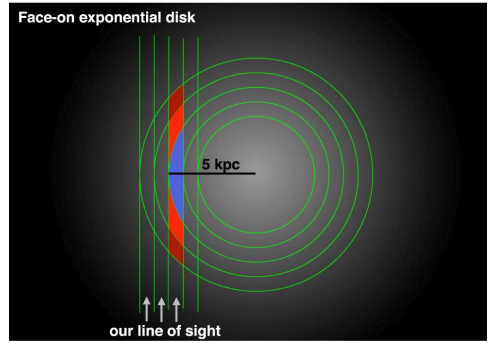} 
   \caption{This figure illustrates that when we observe an edge-on galaxy, we observe not only light from the annulus (green circles) at the projected galactocentric radius (blue), but also light from larger galactocentric radii that happens to be along the line of sight.  For the apertures measured in Section \ref{strippingrate}, we must model the light distribution in the outer disk in order to determine how much of the observed flux is actually coming from the radius we are interested in.}
   \label{diskcontam}
\end{figure}
In this section we discuss corrections for the three effects identified in Section \ref{strippingrate} which complicate the interpretation of the UV-optical colors used to derive stellar population ages.  Optical \textbf{dust extinction} is observed on the trailing side, which likely reddens the observed colors and cause the stellar populations to appear older than they actually are.  The \textbf{composite stellar population along the line of sight} is a consequence of observing a nearly edge-on galaxy - within a single aperture, one measures stars at many galactocentric radii.  If there is a stellar population gradient (as there is on the leading side of NGC 4330), one observes a composite stellar population, and it may be difficult to uniquely determine the star formation history as a function of galactocentric radius.  Contamination from stars in the outer disk leads to redder or "older" colors being measured at a given radius.  Finally, \textbf{asymmetrical quenching times} arise because stars now at the leading side were not always at the leading side, and may not have formed on the leading side, because the galaxy rotates.  For example, any stars 150 - 300 Myr old which are now on the leading side between 5 - 8 kpc formed on the trailing side, since 200 Myr is half of the galaxy's rotational period.  Some of the stars now on the leading side reflect the conditions on the trailing side in the past.

\begin{enumerate}
\item \textbf{Dust Extinction.}

\indent Figure \ref{sspgrid} includes a reddening vector calculated for A$_{FUV}$=1, which corresponds to E(B-V)=0.035 using the wavelength-dependent dust extinction formulae of Calzetti (2001).  This is a modest amount of reddening - for a location near R$_{25}$ of a non-stripped late-type spiral, Mu{\~n}oz-Mateos et al.(2009) find that A$_{FUV}$=2.   In particular, there is neither optical dust extinction nor mid-far IR dust emission on the leading side beyond 5 kpc (Abramson et al. 2009).  NGC 4330 has had its outer disk stripped and now lacks any ISM tracers on the leading side.  As a result of stripping, the outer disk probably has much less dust extinction than a typical spiral.

\item \textbf{Composite Stellar Population Along the Line of Sight.}

\indent Because we view the galaxy nearly edge-on, each aperture measured in Section \ref{strippingrate} contains not only light from the projected galactocentric radius, but also light from further out in the disk which happens to be along our line of sight (Figure \ref{diskcontam}).  To correct for this effect, we model the radial light distribution of the outer disk and subtract the amount of light in each band which comes from a radial bin further out in the disk than the one we are interested.   We used Galfit (Peng et al. 2002) to fit an exponential disk to the galaxy's light distribution at each wavelegth, with separate fits to the upturn and tail sides.  By doing this, we can estimate what fraction of the flux in each aperture is actually coming from the radial bin that we are interested in.  The different scale lengths of the UV and r-band disks result in different correction factors for each wavelength, changing the FUV-NUV and NUV-r colors.  

\indent We also tried fitting a power-law function, and a combination of exponential and power-law functions, and found that every function that provided a very good match to the data produced similar correction factors.  However, if the actual disk is non-axisymmetric, which is likely, the correction factors could be somewhat different.  There are several possible ways to fit each light distribution, and the derived functions are most affected by which portion of the disk's radial light distribution is fit.  Although it is difficult to uniquely determine the star formation history at a given galactocentric radius, we can use this technique to estimate how much the observed colors might change if we were able to remove the light from the outer disk, which allows us to estimate the range of possible quenching times for the outer disk.  

\indent The range of possible values resulting from the outer disk correction are given by the colored boxes in Figure \ref{sspgrid}.  Without applying the correction, the derived stellar population quenching times (Figure \ref{sspgrid}, filled symbols) are quite old (250 - 400 Myr), even those at 5 kpc, just beyond the upturn structure at the leading edge.  Three H$\alpha$-bright regions at the edge of the upturn structure have uncorrected quenching times of between 50 and 200 Myr, although the lifetime of an HII region is generally closer to $\sim$3 Myr.  After applying the correction (the parameter space within the open boxes in Figure \ref{sspgrid}), the point closest to the upturn at 5 kpc may have a quenching time of $\ll$50 Myr, which is much more realistic.  The correction introduces additional error which is hard to estimate, but it provides a general idea of how much the outer disk may redden the colors in the apertures we observe.

 \begin{figure*}[htbp] 
    \centering
    \includegraphics[width=7in]{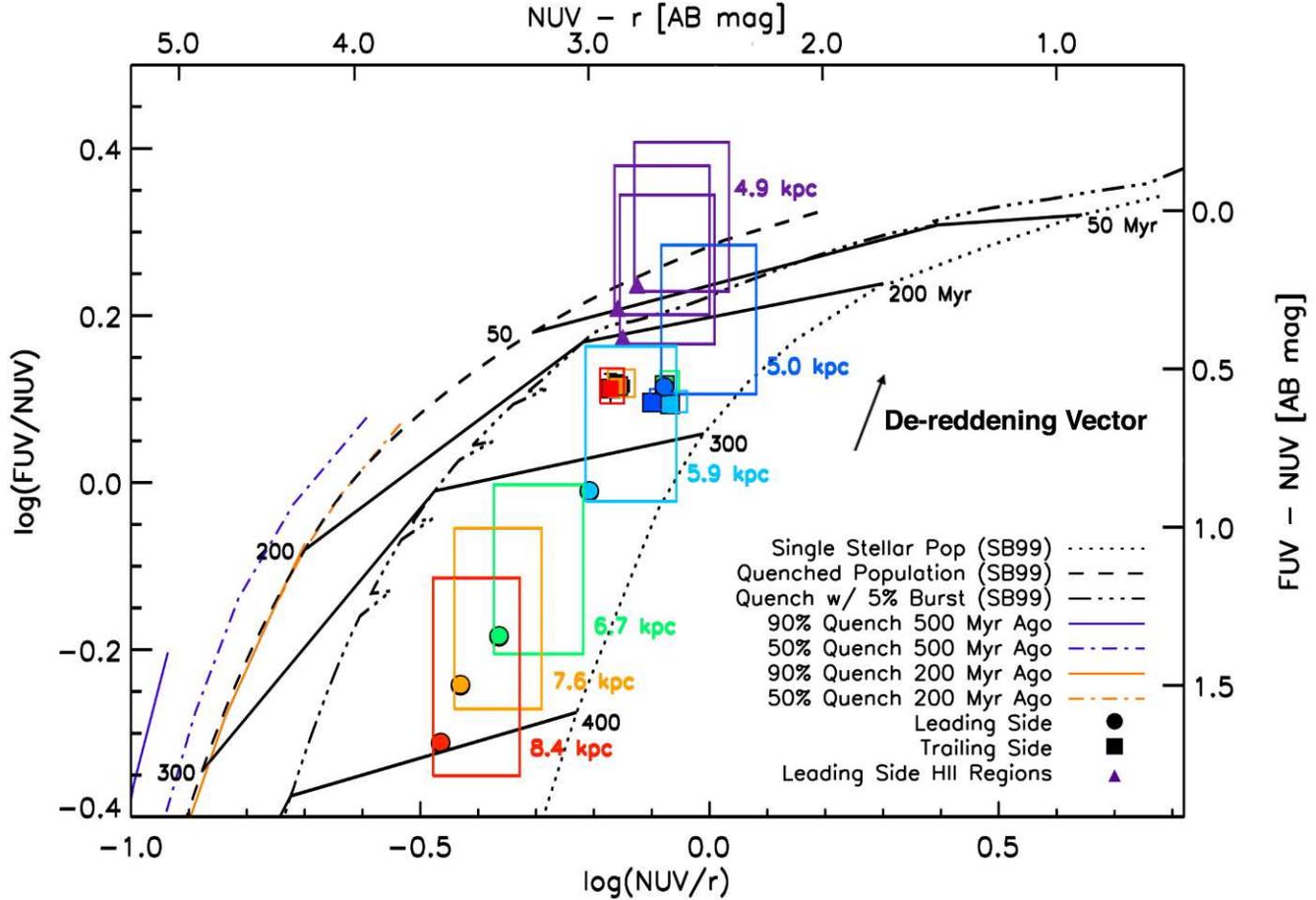} 
    \caption{Same as Figure \ref{sspgrid}, but with colored lines indicating partially-quenched star formation histories.  The solid lines indicate stellar populations whose star formation rates were quenched by 90\% either 500 Myr (blue) or 200 Myr (orange) ago, and had the remaining star formation quenched more recently.  The dashed lines indicate a 50\% quench at 500 Myr (blue) and 200 Myr (orange) ago.}  
    \label{ssppartialquench}
 \end{figure*}
 
 \item \textbf{Asymmetrical Quenching Times.}  
 
\indent The quenching time as a function of position does not have a simple relationship with the star formation history as a function of position because stars move.  Stars now at the leading side were not always at the leading side, and may not have formed on the leading side, because the galaxy rotates.  For example, some of the stars now on the leading side reflect the conditions on the trailing side in the past: any stars 150 - 300 Myr old which are now on the leading side at radii between 5 - 8 kpc formed on the trailing side, since 200 Myr is half of the galaxy's rotational period.  This leads to additional uncertainty in the stripping rates we derive.  NGC 4330's maximum HI rotational speed is 130 km/s, so for a radius of 5 - 8 kpc, the rotational period is 230 - 380 Myr.  This is comparable to the quenching time gradient on what currently constitutes the leading side, $\sim$300 Myr, so the stellar populations we observe at the leading side of the galaxy were not simply quenched in situ at their current location.

\indent In order to approximate a gradual quenching of star formation, as might happen if a given stellar population suffered somewhat suppressed star formation when the ISM was partially removed at the start of stripping, then kept evolving and was fully quenched on the leading side, we have used models with two episodes of quenching.  First, the star formation rate was decreased by 50\% or 90\% either 200 or 500 Myr ago.  Then the star formation was completely quenched at a range of more recent times -
the results are the colored lines in Figures \ref{ssppartialquench}.  The highest log(FUV/NUV) value for each line represents a quench of the remaining star formation 50 Myr ago, with the second quench happening progressively longer ago.  For example, the solid orange line represents a 90\% quench 200 Myr ago, and if the remaining 10\% of star formation was quenched 50 Myr ago, the colors are very similar to a full quench 200 Myr ago.  

\indent Interestingly, the two-quench models are not a good fit to the observed UV/optical colors - they are a worse fit than the quench plus burst model.   

\end{enumerate}


\begin{thebibliography}{99}
\bibitem[Abadi et al.(1999)]{aba99} Abadi, M. G., Moore, B., \& Bower, R. G. 1999, \mnras, 308, 947
\bibitem[Abramson \& Kenney(2009)]{2009eimw.confE..15A} Abramson, A., \& Kenney, J.~D.~P.\ 2009, in: K. Sheth, A. Noriega- 
Crespo, J. Ingalls, R. Paladini (eds.), \emph{The Evolving ISM in the Milky Way and Nearby Galaxies}, online at http://ssc.spitzer.caltech.edu/mtgs/ismevol/, E15
\bibitem[Balogh et al.(1998)]{1998ApJ...504L..75B} Balogh, M.~L., Schade, D., Morris, S.~L., Yee, H.~K.~C., Carlberg, R.~G., \& Ellingson, E.\ 1998, \apjl, 504, L75 
\bibitem[Bekki(1999)] {bek99} Bekki, K. 1999, \apj, 510, L15
\bibitem[ B{\"o}hringer et al.(1994)] {1994Natur.368..828B} B{\"o}hringer, H., Briel, U.~G., Schwarz, R.~A., Voges, W., Hartner, G., 
\& Tr{\"u}mper, J.\ 1994, \nat, 368, 828 
\bibitem[Butcher \& Oemler(1978)] {bo78} Butcher, H., \& Oemler, A., Jr. 1978, \apj, 226, 559
\bibitem[Butcher \& Oemler(1984)] {bo84} Butcher, H., \& Oemler, A., Jr. 1984, \apj, 285, 426
\bibitem[Calzetti(2001)]{2001NewAR..45..601C} Calzetti, D.\ 2001, New Astronomy Review, 45, 601 
\bibitem[Chung et al.(2007)] {tails} Chung, A., van Gorkom, J. H., Kenney, J. D. P., \& Vollmer, B. 2007, \apj, 659, L115
\bibitem[Chung et al.(2009)]{2009AJ....138.1741C} Chung, A., van Gorkom, 
J.~H., Kenney, J.~D.~P., Crowl, H., \& Vollmer, B.\ 2009, \aj, 138, 1741 
\bibitem[Conroy et al.(2009)]{2009ApJ...699..486C} Conroy, C., Gunn, J.~E., \& White, M.\ 2009, \apj, 699, 486 
\bibitem[Conroy et al.(2010)]{2010ApJ...708...58C} Conroy, C., White, M., \& Gunn, J.~E.\ 2010, \apj, 708, 58 
\bibitem[Conroy \& Gunn(2009)]{2009arXiv0911.3151C} Conroy, C., \& Gunn, J.~E.\ 2009, arXiv:0911.3151 
\bibitem[Cortese et al.(2007)]{cor} Cortese, L.; Marcillac, D.; Richard, J.; Bravo-Alfaro, H.; Kneib, J.-P.; Rieke, G.; Covone, G.; Egami, E.; Rigby, J.; Czoske, O.; Davies, J., 2007, \mnras, 376, 157
\bibitem[Crowl \& Kenney(2008)]{crowl08} Crowl, H. H.; Kenney, J. D. P., 2008, \aj, 136, 1623
\bibitem[Crowl et al.(2005)] {crowl4402} Crowl, H. H., Kenney, J. D. P., van Gorkom, J. H., \& Vollmer, B. 2005, \aj, 130, 65
\bibitem[Dressler(1980)] {d1980} Dressler, A. 1980, \apj, 236, 351
\bibitem[Dressler et al.(1997)] {d1997} Dressler, A., Oemler, A., Couch, W. J., Smail, I., Ellis, R. S., Barger, A., Butcher, H., Poggianti, B. M., \& Sharples, R. M. 1997, \apj, 490, 577
\bibitem[Gunn \& Gott(1972)] {gg72} Gunn, J. E. \& Gott, J. R.  1972, \apj, 176, 1
\bibitem[Haynes \& Giovanelli(1984)]{1984AJ.....89..758H} Haynes, M.~P., \& Giovanelli, R.\ 1984, \aj, 89, 758 
\bibitem[Haynes \& Giovanelli(1986)] {hg} Haynes, M. P. \& Giovanelli, R. 1986, \apj, 306, 466
\bibitem[J{\'a}chym et al.(2009)]{2009A&A...500..693J} J{\'a}chym, P., K{\"o}ppen, J., Palou{\v s}, J., \& Combes, F.\ 2009, \aap, 500, 693 
\bibitem[Kenney et al.(2010)] {k4522-hst} Kenney et al. 2010, in preparation 
\bibitem[Kenney et al.(2004)] {kvgv4522} Kenney, J. D. P., van Gorkom, J. H., \& Vollmer, B. 2004, \aj, 127, 3361\bibitem[] {kk04} Koopmann, R. A. \& Kenney, J. D. P. 2004, \apj, 613, 851
\bibitem[Kenney et al.(2008)] {2008ApJ...687L..69K} Kenney, J.~D.~P., Tal, T., Crowl, H.~H., Feldmeier, J., \& Jacoby, G.~H.\ 2008, \apjl, 687, L69 
\bibitem[Kinney et al.(2000)]{2000ApJ...537..152K} Kinney, A.~L., Schmitt, H.~R., Clarke, C.~J., Pringle, J.~E., Ulvestad, J.~S., \& Antonucci, R.~R.~J.\ 2000, \apj, 537, 152 
\bibitem[Koopmann \& Kenney(2004)] {kk04b} Koopmann, R. A. \& Kenney, J. D. P. 2004, \apj, 613, 866
\bibitem[Kronberger et al.(2008)]{2008A&A...481..337K} Kronberger, T., Kapferer, W., Ferrari, C., Unterguggenberger, S., \& Schindler, S.\ 2008, \aap, 481, 337 
\bibitem[Krumholz \& McKee(2008)]{km08} Krumholz, M. R.; McKee, C. F., 2008, Nature, 451, 1082
\bibitem[Larson et al.(1980)] {larson} Larson, R. B., Tinsley, B. M., \& Caldwell, C. 1980, \apj, 237, 692
\bibitem[Lee et al.(2009)]{2009ApJ...706..599L} Lee, J.~C., et al.\ 2009, \apj, 706, 599 
\bibitem[Leitherer et al.(1999)]{star99} Leitherer et al. 1999, \apjs, 123, 3L
\bibitem[Martin et al.(2005)]{galex} Martin et al. 2005, \apjl, 619L,1M
\bibitem[Moore et al.(1998)]{moore98} Moore, B.; Lake, G.; Katz, N., 1998, \apj, 495, 139
\bibitem[Moore et al.(1999)]{moore99} Moore, B.; Lake, G.; Quinn, T.; Stadel, J., 1999, \mnras, 304, 465
\bibitem[Moran et al.(2007)] {m-aph} Moran, S. M., Ellis, R. S., Treu, T., Smith, G. P., Rich, R. M., \& Smail, I. 2007, \apj, 671, 1503
\bibitem[Morrissey et al.(2007)]{morrissey} Morrissey, P., et al. 2007, \apjs, 173, 682
\bibitem[Mu{\~n}oz-Mateos et al.(2009)]{2009ApJ...701.1965M} Mu{\~n}oz-Mateos, J.~C., et al.\ 2009, \apj, 701, 1965 
\bibitem[Murphy et al.(2008)]{Murph08} Murphy, E. J.; Helou, G.; Kenney, J. D. P.; Armus, L.; Braun, R., 2008, \apj, 678, 828
\bibitem[Murphy et al.(2009)]{Murph09} Murphy, E. J.; Kenney, J. D. P.; Helou, G.; Chung, A.; Howell, J. H., 2009, \apj, 694, 1435
\bibitem[Nulsen(1982)] {nul82} Nulsen, P. E. J. 1982, \mnras, 198, 1007
\bibitem[Oey \& Kennicutt(1997)]{oey97} Oey, M. S.; Kennicutt, R. C., Jr., 1997, \mnras, 291, 827
\bibitem[Owen et al.(1974)]{owen} Owen, F. N.; Keel, W. C.; Wang, Q. D.; Ledlow, M. J.; Morrison, G. E., 2006, \aj, 131, 1974
\bibitem[Peng et al.(2002)]{galfit}Peng, C. Y., Ho, L. C., Impey, C. D., \& Rix, H. W. 2002, \aj, 124, 266
\bibitem[Poggianti et al.(1999)] {pog99} Poggianti, B. M., Smail, I., Dressler, A., Couch, W. J., Barger, A. J., Butcher, H., Ellis, R. S., \& Oemler, A., Jr. 1999, \apj, 518, 576
\bibitem[Poggianti et al.(2008)]{pog2009}Poggianti, B. M., et al., 2009, \apj, 693, 112
\bibitem[Quillis et al.(2000)]{2000Sci...288.1617Q} Quilis, V., Moore, B., \& Bower, R.\ 2000, Science, 288, 1617 
\bibitem[Roediger(2009)]{2009AN....330..888R} Roediger, E.\ 2009, Astronomische Nachrichten, 330, 888 
\bibitem[Roediger \& Br{\"u}ggen(2006)]{r06} Roediger, E. \& Br{\"u}ggen, M., 2006, \mnras, 369, 567
\bibitem[Roediger \& Br{\"u}ggen(2007)]{ro07} Roediger, E. \& Br{\"u}ggen, M., 2007, \mnras, 380, 1399
\bibitem[Roediger \& Br{\"u}ggen(2008)]{2008MNRAS.388L..89R} Roediger, E., \& Br{\"u}ggen, M.\ 2008, \mnras, 388, L89 
\bibitem[Roediger \& Br{\"u}ggen(2008]{2008MNRAS.388..465R} Roediger, E., \& Br{\"u}ggen, M.\ 2008, \mnras, 388, 465 
\bibitem[Schulz \& Struck(2001)]{2001MNRAS.328..185S} Schulz, S., \& Struck, C.\ 2001, \mnras, 328, 185 
\bibitem[Solanes et al.(2001)]{2001ApJ...548...97S} Solanes, J.~M., 
Manrique, A., Garc{\'{\i}}a-G{\'o}mez, C., Gonz{\'a}lez-Casado, G., 
Giovanelli, R., \& Haynes, M.~P.\ 2001, \apj, 548, 97 
\bibitem[Sun \& Vikhlinin(2005)]{sun05} Sun, M. \& Vikhlinin, A. 2005, \aj, 621, 718
\bibitem[Sun et al.(2007)]{2007ApJ...671..190S} Sun, M., Donahue, M., 
\& Voit, G.~M.\ 2007, \apj, 671, 190 
\bibitem[Thilker et al.(2005)]{2005ApJ...619L..67T} Thilker, D.~A., et al.\ 2005, \apjl, 619, L67 
\bibitem[Thilker et al.(2007)]{thilk07} Thilker, D. et al., 2007, \apjs, 173, 538
\bibitem[Tonnesen \& Bryan(2009)]{2009ApJ...694..789T} Tonnesen, S., \& Bryan, G.~L.\ 2009, \apj, 694, 789 
\bibitem[Tonnesen \& Bryan(2010)]{2010ApJ...709.1203T} Tonnesen, S., \& Bryan, G.~L.\ 2010, \apj, 709, 1203 
\bibitem[Tuffs et al.(2004)]{2004A&A...419..821T} Tuffs, R.~J., Popescu, C.~C., V{\"o}lk, H.~J., Kylafis, N.~D., \& Dopita, M.~A.\ 2004, \aap, 419, 821
\bibitem[van den Bergh(1976)]{1976ApJ...206..883V} van den Bergh, S.\ 1976, \apj, 206, 883 
\bibitem[van der Wel et al.(2010)]{2010ApJ...714.1779V} van der Wel, A., Bell, E.~F., Holden, B.~P., Skibba, R.~A., \& Rix, H.-W.\ 2010, \apj, 714, 1779
\bibitem[Vollmer et al.(2001)] {vol01} Vollmer, B., Cayatte, V., Balkowski, C., \& Duschl, W. J. 2001, \apj, 561, 708
\bibitem[Vollmer et al.(2004)]{2004A&A...419...35V} Vollmer, B., Balkowski, C., Cayatte, V., van Driel, W., \& Huchtmeier, W.\ 2004, \aap, 419, 35
\bibitem[Vollmer et al.(2009)]{2009A&A...496..669V} Vollmer, B., Soida, M., Chung, A., Chemin, L., Braine, J., Boselli, A., \& Beck, R.\ 2009, \aap, 496, 669 
\bibitem[Vollmer et al.(2006)]{2006A&A...453..883V} Vollmer, B., Soida, M., Otmianowska-Mazur, K., Kenney, J.~D.~P., van Gorkom, J.~H., \& Beck, R.\ 2006, \aap, 453, 883 
\bibitem[Vollmer(2009)]{2009A&A...502..427V} Vollmer, B.\ 2009, \aap, 502, 427 
\bibitem[Weidner \& Kroupa(2006)]{wc} Weidner, C. \& Kroupa, P., 2006, \mnras, 365, 1333
\bibitem[Yasuda et al.(1997)]{1997ApJS..108..417Y} Yasuda, N., Fukugita, 
M., \& Okamura, S.\ 1997, \apjs, 108, 417 

\end{thebibliography}
\end{document}